\documentclass[trackchanges,twocolumn]{aastex62}
\submitjournal{The Astrophysical Journal}
\usepackage{graphicx}
\usepackage{float}
\usepackage{amsmath}
\usepackage[caption=false]{subfig}
\usepackage[utf8]{inputenc}

\shorttitle{MGRO J1908+063 with HAWC}
\shortauthors{HAWC Collaboration}

\usepackage{lineno}

\usepackage{xcolor}

\begin{document}

\title{HAWC Study of the Ultra-High-Energy Spectrum of MGRO J1908+06}
\author{A.~Albert}
\affiliation{Physics Division, Los Alamos National Laboratory, Los Alamos, NM, USA }
\author{R.~Alfaro}
\affiliation{Instituto de F\'{i}sica, Universidad Nacional Autónoma de México, Ciudad de Mexico, Mexico }
\author{C.~Alvarez}
\affiliation{Universidad Autónoma de Chiapas, Tuxtla Gutiérrez, Chiapas, México}
\author{J.D.~Álvarez}
\affiliation{Universidad Michoacana de San Nicolás de Hidalgo, Morelia, Mexico }
\author{J.R.~Angeles Camacho}
\affiliation{Instituto de F\'{i}sica, Universidad Nacional Autónoma de México, Ciudad de Mexico, Mexico }
\author{J.C.~Arteaga-Velázquez}
\affiliation{Universidad Michoacana de San Nicolás de Hidalgo, Morelia, Mexico }
\author{D.~Avila Rojas}
\affiliation{Instituto de F\'{i}sica, Universidad Nacional Autónoma de México, Ciudad de Mexico, Mexico }
\author{H.A.~Ayala Solares}
\affiliation{Department of Physics, Pennsylvania State University, University Park, PA, USA }
\author{R. ~Babu}
\affiliation{Department of Physics, Michigan Technological University, Houghton, MI, USA }
\author{E.~Belmont-Moreno}
\affiliation{Instituto de F\'{i}sica, Universidad Nacional Autónoma de México, Ciudad de Mexico, Mexico }
\author{C.~Brisbois}
\affiliation{Department of Physics, University of Maryland, College Park, MD, USA }
\author{K.S.~Caballero-Mora}
\affiliation{Universidad Autónoma de Chiapas, Tuxtla Gutiérrez, Chiapas, México}
\author{T.~Capistrán}
\affiliation{Instituto de Astronom\'{i}a, Universidad Nacional Autónoma de México, Ciudad de Mexico, Mexico }
\author{A.~Carramiñana}
\affiliation{Instituto Nacional de Astrof\'{i}sica, Óptica y Electrónica, Puebla, Mexico }
\author{S.~Casanova}
\affiliation{Institute of Nuclear Physics Polish Academy of Sciences, PL-31342 IFJ-PAN, Krakow, Poland }
\author{U.~Cotti}
\affiliation{Universidad Michoacana de San Nicolás de Hidalgo, Morelia, Mexico }
\author{J.~Cotzomi}
\affiliation{Facultad de Ciencias F\'{i}sico Matemáticas, Benemérita Universidad Autónoma de Puebla, Puebla, Mexico }
\author{S.~Coutiño de León}
\affiliation{Department of Physics, University of Wisconsin-Madison, Madison, WI, USA }
\author{E.~De la Fuente}
\affiliation{Departamento de F\'{i}sica, Centro Universitario de Ciencias Exactase Ingenierias, Universidad de Guadalajara, Guadalajara, Mexico }
\author{C.~de León}
\affiliation{Universidad Michoacana de San Nicolás de Hidalgo, Morelia, Mexico }
\author{R.~Diaz Hernandez}
\affiliation{Instituto Nacional de Astrof\'{i}sica, Óptica y Electrónica, Puebla, Mexico }
\author{B.L.~Dingus}
\affiliation{Physics Division, Los Alamos National Laboratory, Los Alamos, NM, USA }
\author{M.A.~DuVernois}
\affiliation{Department of Physics, University of Wisconsin-Madison, Madison, WI, USA }
\author{M.~Durocher}
\affiliation{Physics Division, Los Alamos National Laboratory, Los Alamos, NM, USA }
\author{J.C.~Díaz-Vélez}
\affiliation{Departamento de F\'{i}sica, Centro Universitario de Ciencias Exactase Ingenierias, Universidad de Guadalajara, Guadalajara, Mexico }
\author{K.~Engel}
\affiliation{Department of Physics, University of Maryland, College Park, MD, USA }
\author{C.~Espinoza}
\affiliation{Instituto de F\'{i}sica, Universidad Nacional Autónoma de México, Ciudad de Mexico, Mexico }
\author{K.L.~Fan}
\affiliation{Department of Physics, University of Maryland, College Park, MD, USA }
\author{K.~Fang}
\affiliation{Department of Physics, University of Wisconsin-Madison, Madison, WI, USA }
\author{M.~Fernández Alonso}
\affiliation{Department of Physics, Pennsylvania State University, University Park, PA, USA }
\author{N.~Fraija}
\affiliation{Instituto de Astronom\'{i}a, Universidad Nacional Autónoma de México, Ciudad de Mexico, Mexico }
\author{D.~Garcia}
\affiliation{Instituto de F\'{i}sica, Universidad Nacional Autónoma de México, Ciudad de Mexico, Mexico }
\author{J.A.~García-González}
\affiliation{Tecnologico de Monterrey, Escuela de Ingenier\'{i}a y Ciencias, Ave. Eugenio Garza Sada 2501, Monterrey, N.L., Mexico, 64849}
\author{F.~Garfias}
\affiliation{Instituto de Astronom\'{i}a, Universidad Nacional Autónoma de México, Ciudad de Mexico, Mexico }
\author{G.~Giacinti}
\affiliation{Max-Planck Institute for Nuclear Physics, 69117 Heidelberg, Germany}
\author{H.~Goksu}
\affiliation{Max-Planck Institute for Nuclear Physics, 69117 Heidelberg, Germany}
\author{M.M.~González}
\affiliation{Instituto de Astronom\'{i}a, Universidad Nacional Autónoma de México, Ciudad de Mexico, Mexico }
\author{J.A.~Goodman}
\affiliation{Department of Physics, University of Maryland, College Park, MD, USA }
\author{J.P.~Harding}
\affiliation{Physics Division, Los Alamos National Laboratory, Los Alamos, NM, USA }
\author{J. ~Hinton}
\affiliation{Max-Planck Institute for Nuclear Physics, 69117 Heidelberg, Germany}
\author{B.~Hona}
\affiliation{Department of Physics and Astronomy, University of Utah, Salt Lake City, UT, USA }
\author{D.~Huang}
\affiliation{Department of Physics, Michigan Technological University, Houghton, MI, USA }
\author{F.~Hueyotl-Zahuantitla}
\affiliation{Universidad Autónoma de Chiapas, Tuxtla Gutiérrez, Chiapas, México}
\author{P.~Hüntemeyer}
\affiliation{Department of Physics, Michigan Technological University, Houghton, MI, USA }
\author{A.~Iriarte}
\affiliation{Instituto de Astronom\'{i}a, Universidad Nacional Autónoma de México, Ciudad de Mexico, Mexico }
\author{A.~Jardin-Blicq}
\affiliation{Max-Planck Institute for Nuclear Physics, 69117 Heidelberg, Germany}
\affiliation{Department of Physics, Faculty of Science, Chulalongkorn University, 254
Phayathai Road, Pathumwan, Bangkok 10330, Thailand
National Astronomical Research Institute of Thailand (Public
Organization), Don Kaeo, MaeRim, Chiang Mai 50180, Thailand}
\author{V.~Joshi}
\affiliation{Erlangen Centre for Astroparticle Physics, Friedrich-Alexander-Universit\"at Erlangen-N\"urnberg, Erlangen, Germany}
\author{S.~Kaufmann}
\affiliation{Universidad Politecnica de Pachuca, Pachuca, Hgo, Mexico }
\author{D.~Kieda}
\affiliation{Department of Physics and Astronomy, University of Utah, Salt Lake City, UT, USA }
\author{W.H.~Lee}
\affiliation{Instituto de Astronom\'{i}a, Universidad Nacional Autónoma de México, Ciudad de Mexico, Mexico}
\author{J.~Lee}
\affiliation{University of Seoul, Seoul, Rep. of Korea}
\author{H.~León Vargas}
\affiliation{Instituto de F\'{i}sica, Universidad Nacional Autónoma de México, Ciudad de Mexico, Mexico }
\author{J.T.~Linnemann}
\affiliation{Department of Physics and Astronomy, Michigan State University, East Lansing, MI, USA }
\author{A.L.~Longinotti}
\affiliation{Instituto de Astronom\'{i}a, Universidad Nacional Autónoma de México, Ciudad de Mexico, Mexico }
\author{G.~Luis-Raya}
\affiliation{Universidad Politecnica de Pachuca, Pachuca, Hgo, Mexico }
\author{K.~Malone}
\affiliation{Physics Division, Los Alamos National Laboratory, Los Alamos, NM, USA }
\affiliation{Space Science and Applications Group, Los Alamos National Laboratory, Los Alamos, NM, USA} 
\author{V.~Marandon}
\affiliation{Max-Planck Institute for Nuclear Physics, 69117 Heidelberg, Germany}
\author{O.~Martinez}
\affiliation{Facultad de Ciencias F\'{i}sico Matemáticas, Benemérita Universidad Autónoma de Puebla, Puebla, Mexico }
\author{J.~Martínez-Castro}
\affiliation{Centro de Investigaci\'on en Computaci\'on, Instituto Polit\'ecnico Nacional, M\'exico City, M\'exico.}
\author{J.A.~Matthews}
\affiliation{Dept of Physics and Astronomy, University of New Mexico, Albuquerque, NM, USA }
\author{P.~Miranda-Romagnoli}
\affiliation{Universidad Autónoma del Estado de Hidalgo, Pachuca, Mexico }
\author{J.A.~Morales-Soto}
\affiliation{Universidad Michoacana de San Nicolás de Hidalgo, Morelia, Mexico }
\author{E.~Moreno}
\affiliation{Facultad de Ciencias F\'{i}sico Matemáticas, Benemérita Universidad Autónoma de Puebla, Puebla, Mexico }
\author{M.~Mostafá}
\affiliation{Department of Physics, Pennsylvania State University, University Park, PA, USA }
\author{A.~Nayerhoda}
\affiliation{Institute of Nuclear Physics Polish Academy of Sciences, PL-31342 IFJ-PAN, Krakow, Poland }
\author{L.~Nellen}
\affiliation{Instituto de Ciencias Nucleares, Universidad Nacional Autónoma de Mexico, Ciudad de Mexico, Mexico}
\author{M.~Newbold}
\affiliation{Department of Physics and Astronomy, University of Utah, Salt Lake City, UT, USA }
\author{M.U.~Nisa}
\affiliation{Department of Physics and Astronomy, Michigan State University, East Lansing, MI, USA }
\author{R.~Noriega-Papaqui}
\affiliation{Universidad Autónoma del Estado de Hidalgo, Pachuca, Mexico }
\author{L.~Olivera-Nieto}
\affiliation{Max-Planck Institute for Nuclear Physics, 69117 Heidelberg, Germany}
\author{N.~Omodei}
\affiliation{Department of Physics, Stanford University: Stanford, CA 94305–4060, USA}
\author{A.~Peisker}
\affiliation{Department of Physics and Astronomy, Michigan State University, East Lansing, MI, USA }
\author{Y.~Pérez Araujo}
\affiliation{Instituto de Astronom\'{i}a, Universidad Nacional Autónoma de México, Ciudad de Mexico, Mexico }
\author{E.G.~Pérez-Pérez}
\affiliation{Universidad Politecnica de Pachuca, Pachuca, Hgo, Mexico }
\author{C.D.~Rho}
\affiliation{University of Seoul, Seoul, Rep. of Korea}
\author{D.~Rosa-González}
\affiliation{Instituto Nacional de Astrof\'{i}sica, Óptica y Electrónica, Puebla, Mexico }
\author{H.~Salazar}
\affiliation{Facultad de Ciencias F\'{i}sico Matemáticas, Benemérita Universidad Autónoma de Puebla, Puebla, Mexico }
\author{F.~Salesa Greus}
\affiliation{Institute of Nuclear Physics Polish Academy of Sciences, PL-31342 IFJ-PAN, Krakow, Poland }
\affiliation{Instituto de Física Corpuscular, CSIC, Universitat de València, E-46980, Paterna, Valencia, Spain}
\author{A.~Sandoval}
\affiliation{Instituto de F\'{i}sica, Universidad Nacional Autónoma de México, Ciudad de Mexico, Mexico}
\author{M.~Schneider}
\affiliation{Department of Physics, University of Maryland, College Park, MD, USA }
\author{H. Schoorlemmer}
\affiliation{IMAPP, Radboud University Nijmegen, Nijmegen, The Netherlands}
\affiliation{Max-Planck Institute for Nuclear Physics, 69117 Heidelberg, Germany}
\author{J.~Serna-Franco}
\affiliation{Instituto de F\'{i}sica, Universidad Nacional Autónoma de México, Ciudad de Mexico, Mexico }
\author{A.J.~Smith}
\affiliation{Department of Physics, University of Maryland, College Park, MD, USA }
\author{Y.~Son}
\affiliation{University of Seoul, Seoul, Rep. of Korea}
\author{R.W.~Springer}
\affiliation{Department of Physics and Astronomy, University of Utah, Salt Lake City, UT, USA }
\author{O.~Tibolla}
\affiliation{Universidad Politecnica de Pachuca, Pachuca, Hgo, Mexico }
\author{K.~Tollefson}
\affiliation{Department of Physics and Astronomy, Michigan State University, East Lansing, MI, USA }
\author{I.~Torres}
\affiliation{Instituto Nacional de Astrof\'{i}sica, Óptica y Electrónica, Puebla, Mexico }
\author{R.~Torres-Escobedo}
\affiliation{Tsung-Dao Lee Institute and School of Physics and Astronomy, Shanghai Jiao Tong University, Shanghai, China}
\author{R.~Turner}
\affiliation{Department of Physics, Michigan Technological University, Houghton, MI, USA }
\author{F.~Ureña-Mena}
\affiliation{Instituto Nacional de Astrof\'{i}sica, Óptica y Electrónica, Puebla, Mexico }
\author{L.~Villaseñor}
\affiliation{Facultad de Ciencias F\'{i}sico Matemáticas, Benemérita Universidad Autónoma de Puebla, Puebla, Mexico }
\author{X.~Wang}
\affiliation{Department of Physics, Michigan Technological University, Houghton, MI, USA }
\author{I.J.~Watson}
\affiliation{University of Seoul, Seoul, Rep. of Korea}
\author{E.~Willox}
\affiliation{Department of Physics, University of Maryland, College Park, MD, USA }
\author{A.~Zepeda}
\affiliation{Physics Department, Centro de Investigacion y de Estudios Avanzados del IPN, Mexico City, Mexico }
\author{H.~Zhou}
\affiliation{Tsung-Dao Lee Institute and School of Physics and Astronomy, Shanghai Jiao Tong University, Shanghai, China}

\collaboration{HAWC Collaboration}

\author{M. Breuhaus}
\affiliation{Max-Planck Institute for Nuclear Physics, 69117 Heidelberg, Germany}
\affiliation{Member of the International Max-Planck Research School for Astronomy and Cosmic Physics at the University of Heidelberg (IMPRS-HD), Germany}
\author{H.~Li}
\affiliation{Theoretical Division, Los Alamos National Laboratory, Los Alamos, NM, USA}
\author{H.~Zhang}
\affiliation{New Mexico Consortium, Los Alamos, NM 87544, USA}
\affiliation{Department of Physics and Astronomy, Purdue University, West Lafayette, IN 47907, USA}
\correspondingauthor{Kelly Malone}
\email{kmalone@lanl.gov}
\begin{abstract}
We report TeV gamma-ray observations of the ultra-high-energy source MGRO J1908+06 using data from the High Altitude Water Cherenkov (HAWC) Observatory. This source is one of the highest-energy known gamma-ray sources, with emission extending past 200 TeV. Modeling suggests that the bulk of the TeV gamma-ray emission is leptonic in nature, driven by the energetic radio-faint pulsar PSR J1907+0602. Depending on what assumptions are included in the model, a hadronic component may also be allowed. Using the results of the modeling, we discuss implications for detection prospects by multi-messenger campaigns.
\end{abstract}

\section{Introduction}

\subsection{Previous TeV observations}

MGRO J1908+06, which is in the Galactic plane, was originally discovered by the Milagro detector~\citep{Milagro} in the very high-energy regime (median energy $\sim$20 TeV) and subsequently confirmed by other TeV observatories, including H.E.S.S, VERITAS, and ARGO~\citep{HESS,VERITAS,ARGO}.  Both imaging atmospheric Cherenkov telescopes (IACTs) report that the source is extended and has a fairly hard spectral index ($\leq$ 2.2), while ARGO reports a slightly softer index (2.54$\pm$0.36).  

Emission from this region has recently been detected by the High Altitude Water Cherenkov (HAWC) Observatory and included in the collaboration's third catalog (3HWC) with the source name 3HWC J1908+063~\citep{3hwc}. This emission is centered at (287.05$^{\circ}$, 6.39$^{\circ}$) in the (right ascension, declination) J2000 coordinate system. The 1$\sigma$ statistical uncertainty in this location is 0.06$^{\circ}$.

Recently, both HAWC and LHAASO reported that this region is one of only  a handful emitting above 100~TeV~\citep{HECatalog, lhaaso2}. Even at the highest energies, the source remains extended, with the HAWC collaboration reporting a Gaussian width of 0.52$^{\circ}$ above 56 TeV.

The high-energy emission makes this an intriguing source to study. The cosmic-ray spectrum contains a bump known as the ``knee" around 1 PeV~\citep{pdg2021}. Which Galactic sources are capable of accelerating particles to this energy is still an open question. Cosmic-ray interactions with their environment produce neutron pions, which then, in turn, decay to gamma rays. These gamma rays are approximately one order of magnitude less in energy compared to the primary cosmic ray. Therefore, in order to probe the knee of the cosmic ray spectrum studies of 100 TeV gamma rays are essential.  100 TeV gamma rays can also be produced in other manners, such as inverse Compton scattering. In fact, leptonic mechanisms likely dominate the highest-energy sky~\citep{breuhaus, Breuhaus_et_al_2021_LHAASO,2021uhe,2021sudoh}

\subsection{Multi-wavelength/multi-messenger observations}

There are several objects in the region which could serve as counterparts to the TeV emission, including a supernova remnant (SNR) and several pulsars. The radio SNR G40.5-0.5~\citep{GreenSNR} has an estimated age of 20-40 kyrs~\citep{Downes}. The distance to this supernova remnant is quite uncertain, with distances between 3.4 kpc and 8.7 kpc appearing in the literature~\citep{radio}. The SNR is offset from the TeV emission seen by VERITAS, which led the collaboration to conclude that it is not the main source of the TeV emission~\citep{VERITAS}. The SNR has an angular separation of 0.29$^{\circ}$ from the center of the 3HWC source.  

There are also three pulsars nearby. The most energetic is PSR J1907+0602, a radio-faint pulsar that was discovered in the GeV energy range by the Fermi Large Area Telescope (Fermi-LAT)~\citep{Fermi}. According to the ATNF pulsar database\footnote{version 1.63, \url{https://www.atnf.csiro.au/research/pulsar/psrcat/}}~\citep{Manchester2005}, this pulsar has a high spin-down power ($\dot{E}$) of 2.8$\times$10$^{36}$ erg/s, a characteristic age of 19.5 kyr, and is estimated to be 2.37 kpc from the Earth. 

While it is possible that this pulsar could have been born in SNR G40.5-0.5, it seems somewhat unlikely. Given the present distance between them ($\sim$28 pc, assuming both objects are the same distance from the Earth), a transverse velocity 3 times higher than typical due to a supernova kick is required~\citep{VERITAS}.

There has been X-ray emission observed near PSR J1907+0602 with the Chandra X-ray Observatory~\citep{Fermi}. Originally it was thought that this emission may be slightly extended and could be the X-ray pulsar wind nebula (PWN), but recently ~\cite{fermiPWN} have shown that the X-ray data is consistent with a point source and can be attributed to the pulsar. Additionally, a study using XMM-Newton data found no diffuse or extended emission in a 0.75 by 0.75 degree area~\citep{XMM} around the TeV source.

A second pulsar in the region, PSR J1907+0631, is not as energetic as PSR J1907+0602 ($\dot{E}$ of 5.3$\times$10$^{35}$ erg/s)\footnote{version 1.63, \url{https://www.atnf.csiro.au/research/pulsar/psrcat}}. Its location is very close to the center of the SNR G40.5-0.5 and its estimated age is consistent with the estimated age of the SNR. It has been suggested that this is the pulsar formed in the supernova event~\citep{PAFLA}.  

\begin{figure}
\centering
\includegraphics[width=0.5\textwidth]{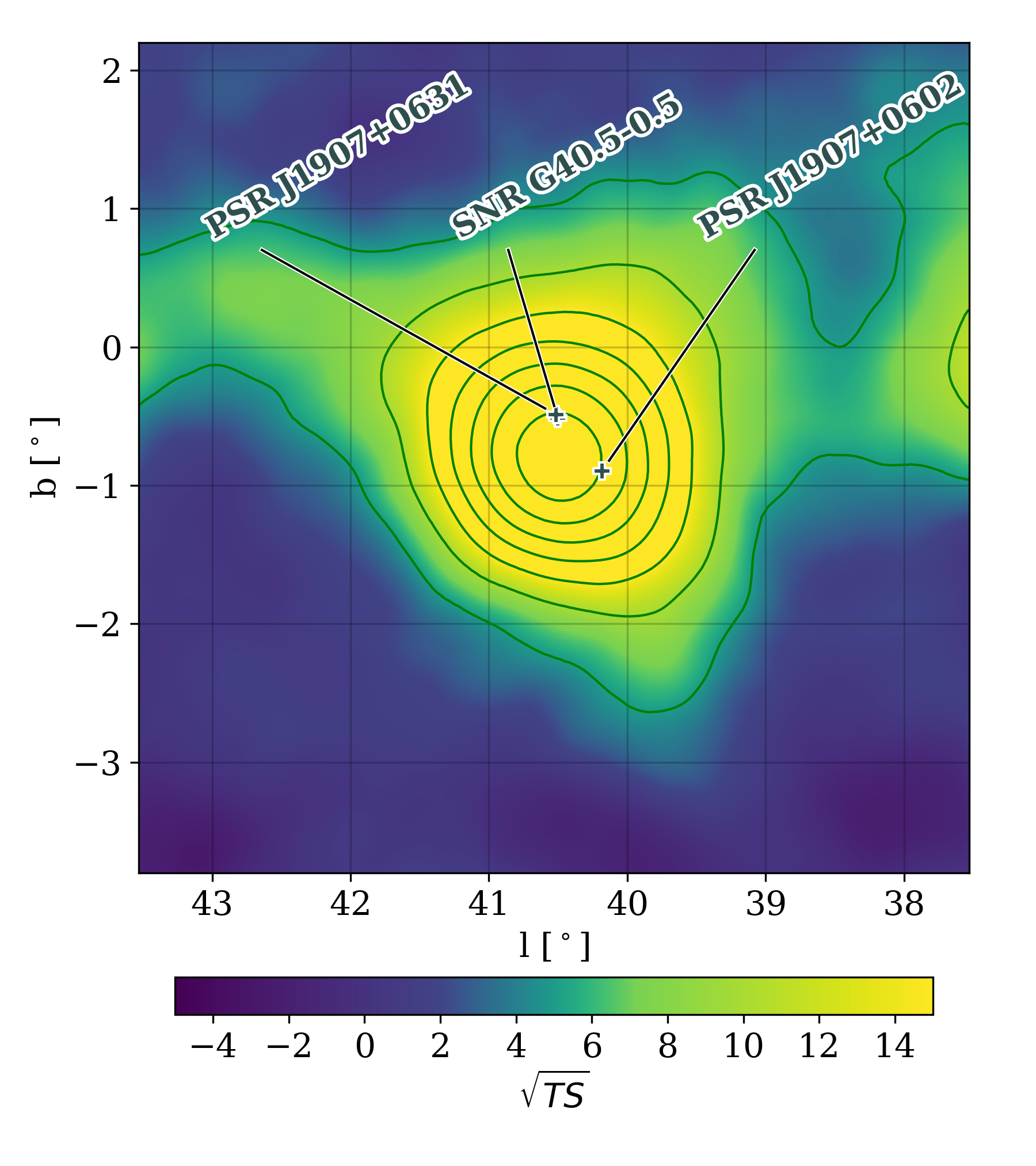}
\caption{HAWC significance map of the region, in Galactic coordinates, with the two pulsars and the SNR labeled. PSR J1907+0631 and SNR G40.5-0.5 are only 0.03$^{\circ}$ away from each other so their markers on this plot overlap. The maximum significance is 38.82$\sigma$. The contours are the 5, 10, 15, 20, 25, 30, and 35$\sigma$ significance contour levels.}
\label{fig:assocations}
\end{figure}

The centroid of the HAWC emission lies roughly in between these two pulsars, roughly $\sim$0.3$^{\circ}$ from both of them. Figure \ref{fig:assocations} shows the HAWC significance map of the region with the possible counterparts to the emission labeled. 

There is also a third pulsar in the region, PSR J1905+0600, but it is weaker ($\dot{E}$ = 5.1$\times$10$^{32}$ erg/s) and older (6 Myr) than the other two pulsars~\citep{Parkes} so it is not expected to contribute to the observed TeV emission. This pulsar is $\sim$0.87$^{\circ}$ away from the centroid of the HAWC emission. Its distance is 8.8 kpc away from the Earth.

If any of the emission is hadronic in nature, neutrinos are also expected. Due to the size and hard spectrum of the TeV gamma-ray emission, MGRO J1908+06 has been frequently suggested as a target for neutrino searches~\citep{gonzalezgarcia,Halzen2017}. In an IceCube search for point-like sources in the astrophysical muon neutrino flux, this region had the second-best $p$-value (highest for a Galactic source), although still consistent with background~\citep{IceCube}. 

A radio study of the region~\citep{radio} resulted in the discovery of new molecular clouds to the eastern, southern and western borders of the radio shell of the SNR. The authors of that paper hypothesized that the TeV emission actually consists of two separate components: a leptonic component powered by PSR J1907+0602 and a hadronic component produced by interactions between G40.5-0.5 and the newly discovered molecular gas. If SNR G40.5-0.5 is 8.5 kpc away, as some distance estimates suggest, it is much further away than PSR J1907+0602 and the source we see may actually consist of two superimposed sources. \cite{2021MNRAS.505.2309C} also suggest that the emission is comprised of two populations. 

Recent observations using Fermi-LAT~\citep{fermiPWN} have resulted in the detection of extended GeV gamma-ray emission in this area, said to be the GeV counterpart of the TeV emission. This emission contains two components: a soft, low-energy ($<$ 10 GeV) component and a harder ($>$ 10 GeV) component. The first component is attributed to molecular clouds surrounding the supernova remnant, while the second is likely leptonic in origin and originates from the PWN of PSR J1908+0602.

\subsection{Description of HAWC and HAWC data}

In this work, we use data from the HAWC Observatory to study 3HWC J1908+063. The HAWC detector consists of 300 water Cherenkov detectors, each instrumented with four photomultiplier tubes. It is designed to detect the byproducts of the extensive air showers that are induced when a gamma ray or a cosmic ray enters the Earth's atmopshere and interacts with particles there. 

Located in the state of Puebla, Mexico, HAWC is sensitive to sources with declinations between -26 and +64 degrees. It is capable of continuously monitoring the sky and has achieved a sensitivity of a few percent of the Crab flux over the last five years~\citep{3hwc}.  More information on the design of HAWC can be found in \cite{Smith2015} and \cite{Crab2017}.

This paper uses a dataset consisting of 1343 days of data collected between June 2015 and June 2019. The data is binned using a 2D scheme of the estimated energy ($\hat{E}$) and the fraction of the HAWC array hit during an air shower event, as described in ~\cite{Crab}. The estimated energy bins are each a quarter-decade in width in log$_{10}$ space; the first bin starts at \mbox{$\hat{E}$ = 1 TeV} and the last bin ends at \mbox{$\hat{E}$ = 316 TeV}. The ``ground parameter" energy estimator is used. This algorithm uses the fit to the lateral distribution function to measure the charge density 40 meters from the shower core, along with the zenith angle of the air shower, to estimate the energy of the primary gamma ray. The standard quality cuts described in ~\cite{Crab} are used.

The paper is organized as follows: In Section \ref{sec:diffusion}, we describe the diffusion model we use to fit data in the 3HWC J1908+063 region. Section \ref{sec:results} gives the best-fit results using this diffusion model. We also compare the results presented here to those obtained by other observatories. A potential spectral hardening feature at the highest energies is also discussed. In Section \ref{sec:modeling}, we discuss possible models to describe the TeV emission from HAWC. In Section \ref{sec:multi}, we discuss implications of this model for detection by observatories operating at different wavelengths and with different messengers. In Section \ref{sec:conclusions} we present the conclusions. 

\section{Description of the diffusion model}\label{sec:diffusion}

The model we fit to the region contains three sources: 3HWC J1908+063 as well as the east and west lobes of SS433. The lobes of SS433 overlap the edge of the significant 3HWC J1908+063 emission. 

Both lobes of SS433 are modeled as point sources with their locations fixed to the reported location in ~\cite{ss433}. As in that paper, they are assumed to emit according to power-law spectra with spectral indices fixed at 2.0:
\begin{equation}\label{eq:ss433Eq}
\frac{dN}{dE} = \phi_0 \left(\frac{E}{\text{20 TeV}}\right)^{-2.0}.
\end{equation} 
The spectral indicies are fixed as it is not possible to fit them due to the low number of counts for these sources. This statistical limitation does not have an effect on the fit parameters of 3HWC J1908+063, which is brighter by orders of magnitude.  The normalization of each lobe, $\phi_0$, is allowed to float separately in the fit.

3HWC J1908+063 is modeled as an extended source with the centroid fixed at the location from the 3HWC catalog: (right ascension = 287.05$^{\circ}$, declination = 6.39$^{\circ}$)~\citep{3hwc}. Three spectral shapes are considered: a power-law, a power-law with an exponential cutoff, and a log-parabolic function. The log parabolic function is found to be significantly preferred, using the Bayesian information criterion~\citep{schwarz1978, kass} (BIC), over other spectral shapes: 

\begin{equation}
\frac{dN}{dE} = \phi_0 \left( \frac{E}{\text{10 TeV}} \right)^{-\alpha - \beta \mathrm{ln} (E/\text{10 TeV})}.
\end{equation}

The flux normalization $\phi_0$, the spectral index $\alpha$, and the curvature parameter $\beta$ are all free parameters in the fit. The BIC for this fit is 139459, while the BIC for a power-law fit is 139523 and the BIC for a power-law with an exponential cutoff is 139491. The $\Delta$BIC between this model and the power-law (power-law with an exponential cutoff) is 64 (32).  A $\Delta$BIC value of more than ten implies very strong evidence against the higher BIC~\citep{kass}. 

The source is spatially extended; a diffusion model is chosen to describe the source.  The model is similar to the one used in the HAWC analysis of the Geminga TeV halo~\citep{Geminga}. The pulsars in this region are much younger than Geminga's pulsar, but are old enough that the source could have begun the transition to a TeV halo. If this is the case, the electrons and positrons are expected to be transported via a diffusive mechanism. This diffusion model assumes that electrons and positrons are continually injected from a central point, with isotropic diffusion. Contributions from the cosmic microwave background (CMB), infrared (IR) and optical photon fields are considered, with the same values as ~\cite{Geminga}. The magnetic field is fixed at 3 $\mu$G.

The spatial morphology for this diffusion model is: 
\begin{equation}\label{eq:morphology}
\frac{dN}{d\Omega} =  \frac{1.22}{\pi^{3/2} \theta_d(E)(\theta+0.06\theta_d(E))}\exp(-\theta^2/\theta_d^2(E)),
\end{equation}
where $N$ is the total flux, $E$ is the gamma-ray energy, $\theta$ is the angle from the source, $\Omega$ denotes a solid angle, and $\theta_d$ is the diffusion angle, which is a free parameter in the fit.  It is related to the diffusion radius, $r_d$, by

\begin{equation}\label{eq:diffRad}
\theta_d = \frac{180}{\pi}\frac{r_d}{d_{src}},
\end{equation}
where $d_{src}$ is the distance from the source to the Earth. Then, 
\begin{equation}\label{eq:diffusion}
r_d = 2 \sqrt{D(E_e)t_E},
\end{equation}
where $D$ is the diffusion coefficient for electrons at energy $E_e$ and $t_E$ is the smaller of the injection time and the cooling time. The mean electron energy, $E_e$, can be calculated from the mean gamma-ray energy, $\langle E \rangle$ as follows~\citep{aharonianbook,Geminga}:
\begin{equation}\label{eq:electronEnergy}
\langle E_e \rangle \approx 17 \langle E \rangle^{0.54 + 0.046\mathrm{log}_{10}(\langle E \rangle /\mathrm{TeV})},
\end{equation}

and $D$ is defined as 
\begin{equation}\label{eq:diffusionEnergy}
D(E_e) = D_0 (E_e / \text{10 GeV})^{\delta}.
\end{equation}

In this paper, $\delta$ is fixed to 1/3, motivated by the Kolmogorov turbulence model describing the magnetic field. In the theory of cosmic ray diffusion, the energy dependence is ($2-\gamma$), where $\gamma$ is the shape of the power spectrum. According to Kolmogorov, $\gamma$ is equal to 5/3 ~\citep{kolmogorov, 1999ApJ...520..204G}.  $D(E_e)$ can then be used in Equation \ref{eq:diffusion} to calculate the diffusion radius.
 
The free parameters ($\phi_0$, $\alpha$, $\beta$ and $\theta_d$ for 3HWC J1908+063 and $\phi_0$ for each of the lobes of SS433) are determined simultaneously via a likelihood fit done using the HAL~\citep{HAL}\footnote{HAWC Accelerated Likelihood; \url{https://github.com/threeML/hawc_hal}} plugin to the 3ML (Multi-mission Maximum Likelihood) software~\citep{threeml}\footnote{\url{https://github.com/threeML/threeML}}. A circular region of interest (radius of 3$^{\circ}$) is used. The diffusion model can be found in \textit{astromodels}\footnote{\url{https://github.com/threeML/astromodels}}, a software package that interfaces with 3ML.

\section{Results}\label{sec:results}
In the following sections, we discuss the best-fit results and compare them to those of other detectors.  We also consider systematic uncertainties and discuss a potential spectral hardening feature at the highest energies.  

\subsection{Best-fit results}\label{sec:bestfitresults}

\begin{figure*}
\includegraphics[width=0.5\textwidth]{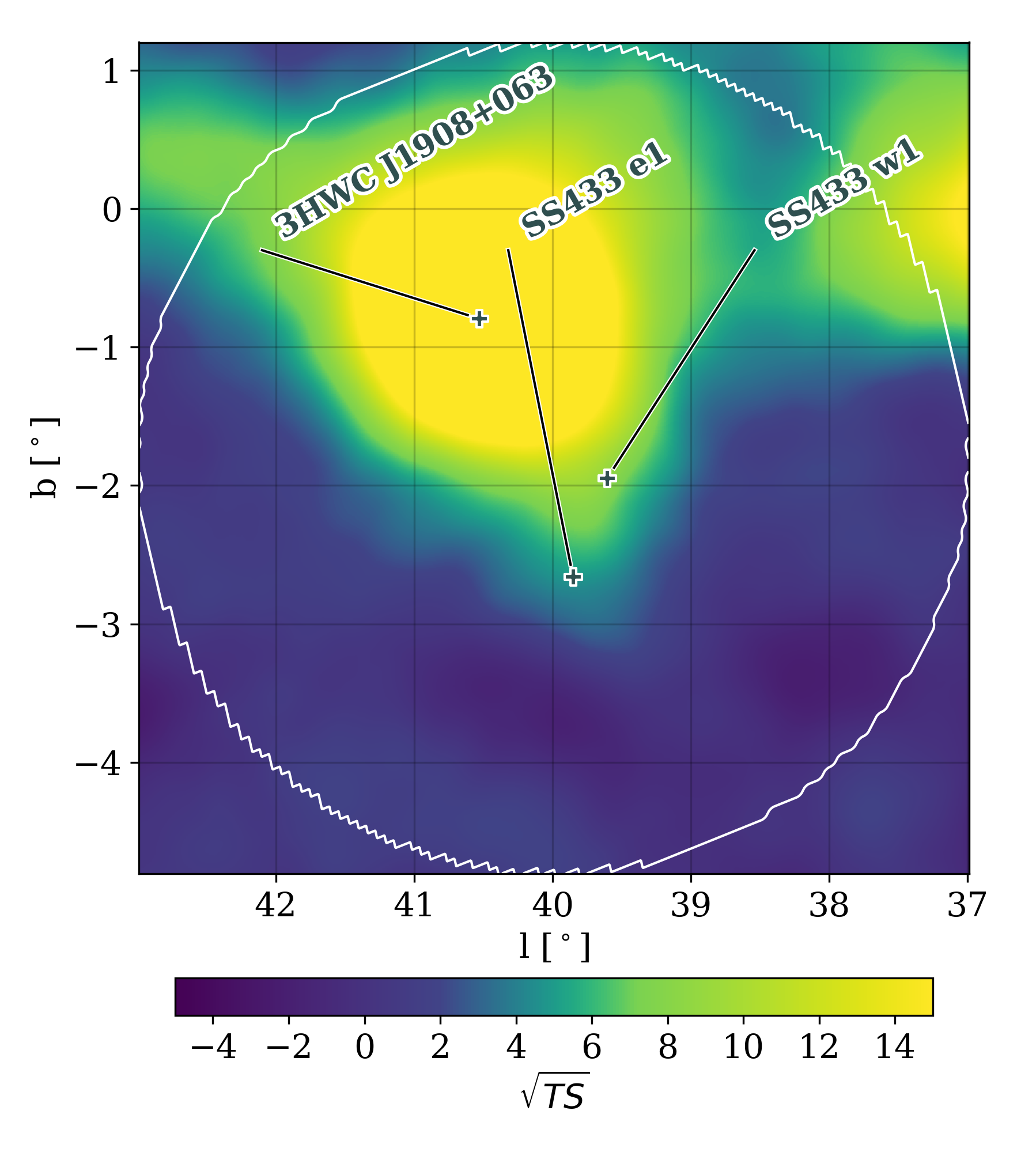}
\subfloat{\includegraphics[width=0.5\textwidth]{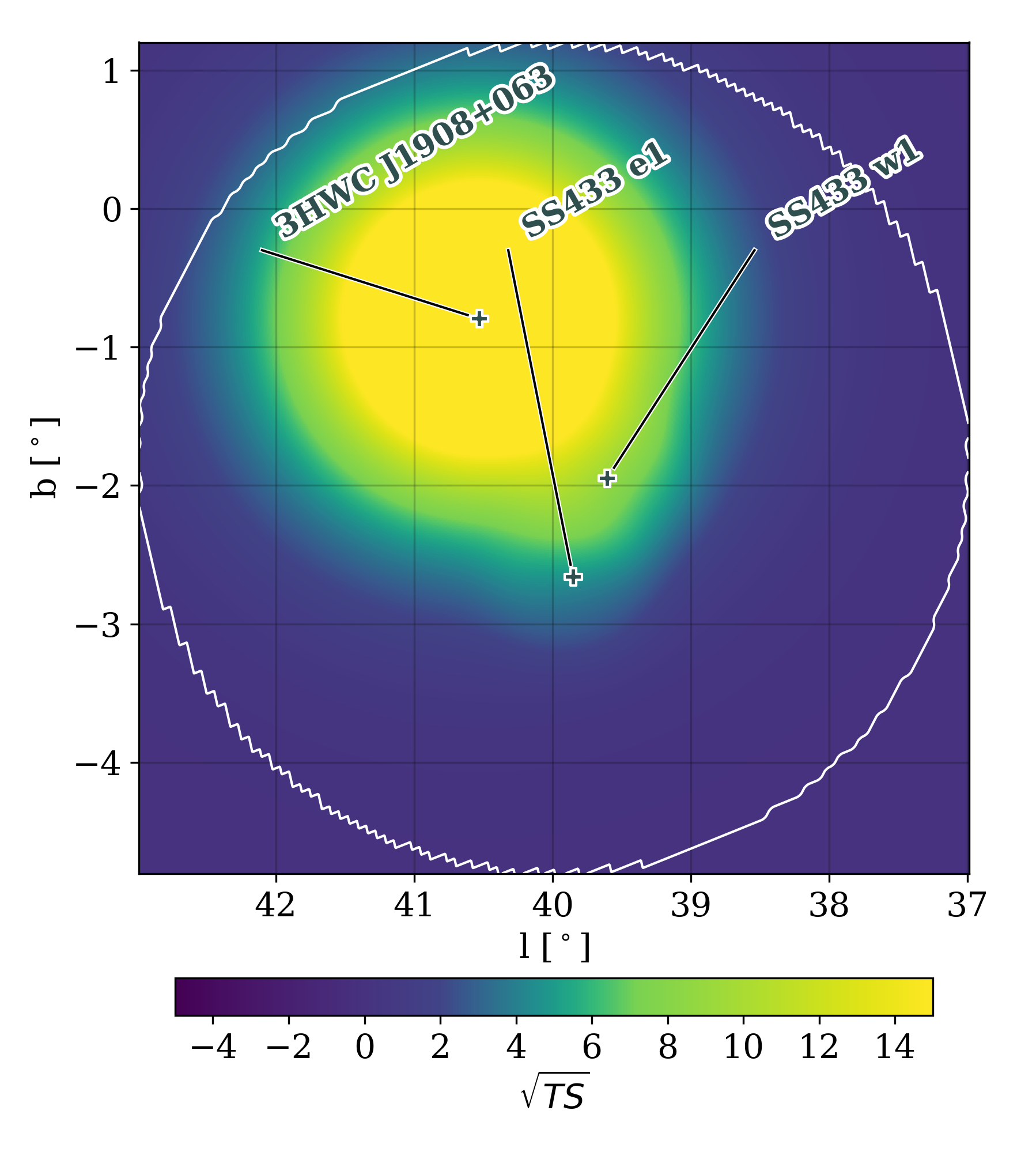}}
\caption{Left: HAWC significance map of the region, for reconstructed energies between 1 and 316 TeV. The map is in Galactic coordinates. For each pixel, the significance is calculated assuming that a disk of radius 0.6$^{\circ}$ is centered at that location and that the spectrum is a power-law with a spectral index of -2.4. Right: The best-fit model used in this analysis. The bulk of the emission comes from 3HWC J1908+063. The east and west lobes of SS433 are modeled as point sources. The maximum significance in this map is 38.82$\sigma$. In both figures, the white circle is the ROI.} 
\label{fig:sigAndModelMap}
\end{figure*}

\begin{figure*}
\includegraphics[width=0.4\textwidth]{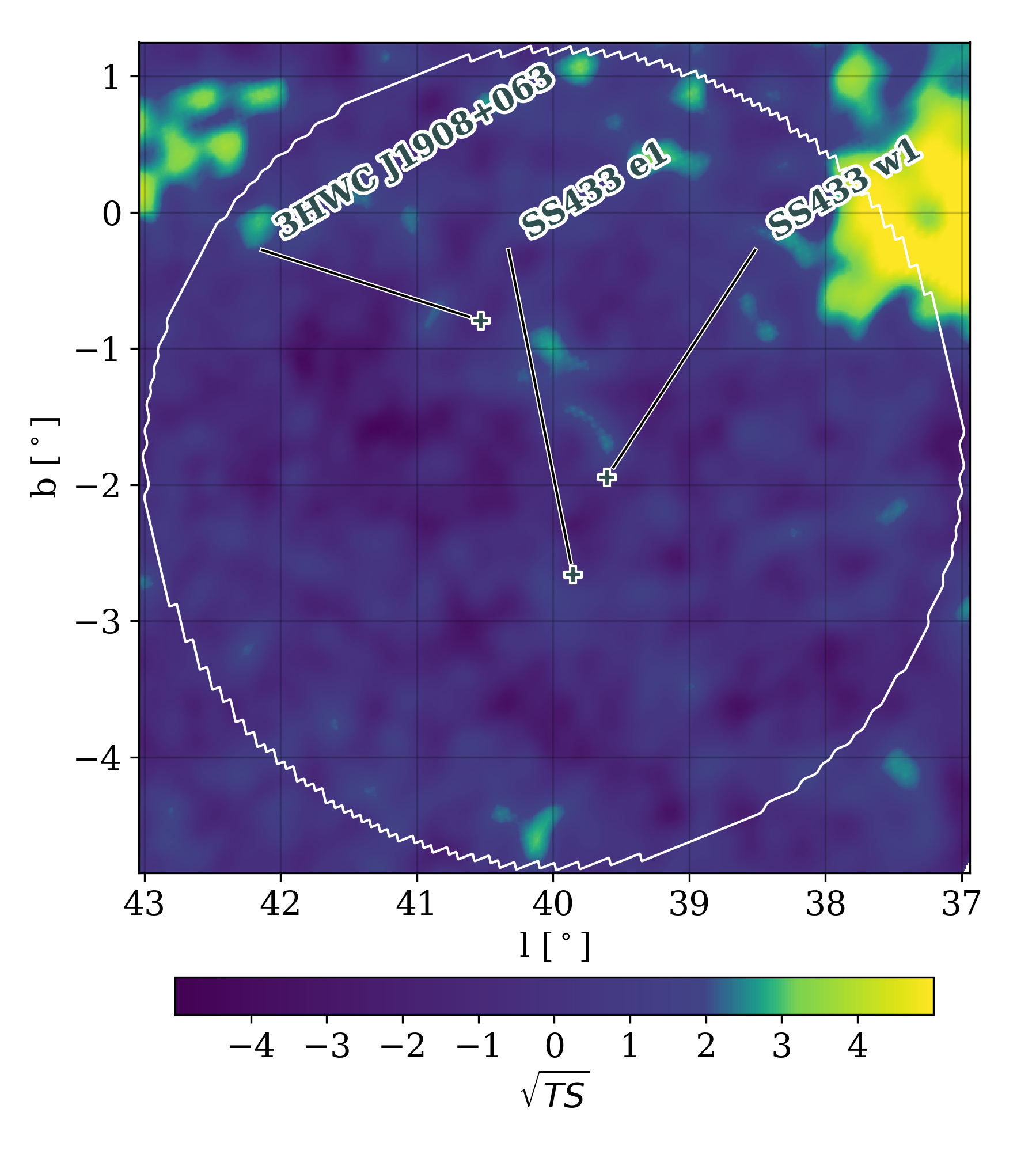}
\subfloat{\includegraphics[width=0.6\textwidth]{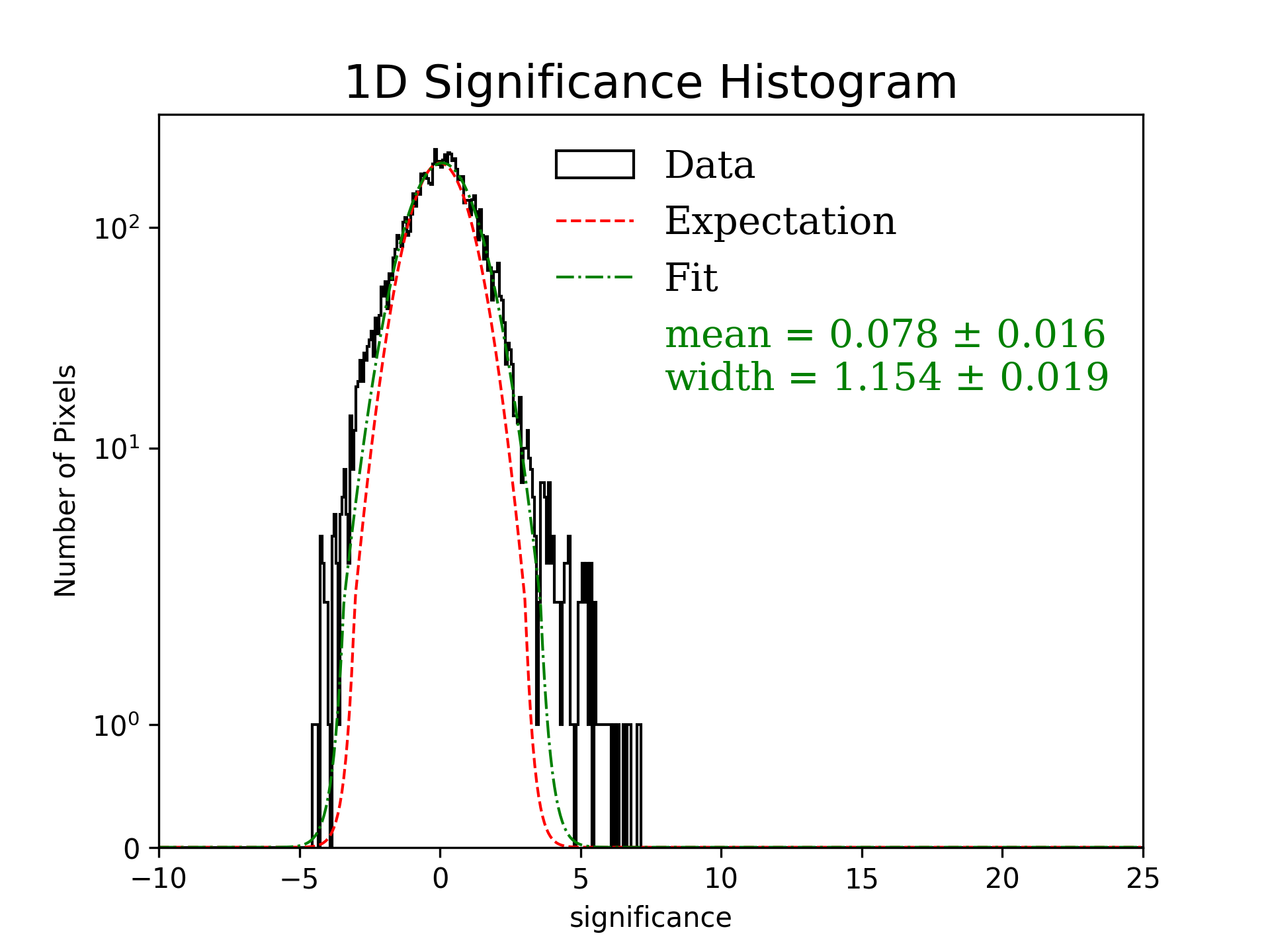}}
\caption{Left: The significance map of the residuals, which is computed by subtracting the best-fit model from the HAWC data. This significance map is computed using a point source morphology to avoid correlations between pixels that are inherently unavoidable in the extended source assumption. The ~$2-3\sigma$ emission in the northwest corner of the ROI could be associated with either PSR J1906+0722 or SNR 3C397. The H.E.S.S detector has presented evidence of emission in this region~\citep{HESS2021ICRC}.   The white circle is the ROI.  Right: The corresponding histogram of the residual values within the region of interest. The excess at $\sigma >$ 5 is due to the emission centered around $l$=37$^{\circ}$, $b$=0$^{\circ}$, the edge of which largely outside the ROI for this analysis. There is no source at this location in the 3HWC catalog and it is likely diffuse emission. See Section \ref{sec:systematics} for a discussion of how diffuse emission affects the results presented here.   } 
\label{fig:resids}
\end{figure*}

 \begin{table*}
\centering
 \begin{tabular}{| c | c | c | c |}
 \hline 
 Parameter & Best-fit value & Statistical uncertainty & Systematic uncertainty \\
 \hline
$\theta_d$ & 1.78$^{\circ}$ & $\pm$ 0.08$^{\circ}$ & $_{-0.28}^{+0.07}$ \\
$\phi_0$ & 1.17 $\times$ 10$^{-13}$ (TeV cm$^{2}$ s)$^{-1}$ &
$\pm$ 0.06 $\times$ 10$^{-13}$ (TeV cm$^{2}$ s)$^{-1}$ & $_{-0.23}^{+0.10}$ $\times$ 10$^{-13}$ (TeV cm$^{2}$ s)$^{-1}$ \\
$\alpha$ & 2.545 & $\pm$ 0.026 & $_{-0.06}^{+0.01}$ \\
$\beta$ & 0.134 & $\pm$ 0.018 & $_{-0.03}^{+0.02}$ \\
$\phi_{SS433E}$ & 2.0 $\times$ 10$^{-16}$ (TeV cm$^{2}$ s)$^{-1}$ & $_{-0.7}^{+1.0}$ $\times$ 10$^{-16}$ (TeV cm$^{2}$ s)$^{-1}$ & $_{-0.1}^{+0.2}$ $\times$ 10$^{-16}$ (TeV cm$^{2}$ s)$^{-1}$  \\
$\phi_{SS433W}$ & 3.0 $\times$ 10$^{-16}$ (TeV cm$^{2}$ s)$^{-1}$ & $_{-0.8}^{+1.1}$ $\times$ 10$^{-16}$ (TeV cm$^{2}$ s)$^{-1}$ & $_{-0.6}^{+0.2}$ $\times$ 10$^{-16}$ (TeV cm$^{2}$ s)$^{-1}$ \\
 \hline 
 \end{tabular}
 \caption{Best-fit values for the continuous injection diffusion model. The first four variables pertain to 3HWC J1908+063 while $\phi_{SS433E}$ and $\phi_{SS433W}$ are the $\phi_0$ values for the East and West lobes of SS433, respectively (see Equation \ref{eq:ss433Eq}). $\theta_d$ is reported at the gamma-ray pivot energy of 10 TeV, which corresponds to electrons with energy $\sim$65 TeV. The column labeled ``Systematic uncertainty" contains the uncertainty from mis-modeling of the detector along with the uncertainty related to modeling of the Galactic diffuse emission. These two uncertainties are combined in quadrature. See Section \ref{sec:systematics} for a discussion of systematic uncertainties. } \label{tab:diffRes} 
 \end{table*}

\begin{figure}
\includegraphics[width=0.5\textwidth]{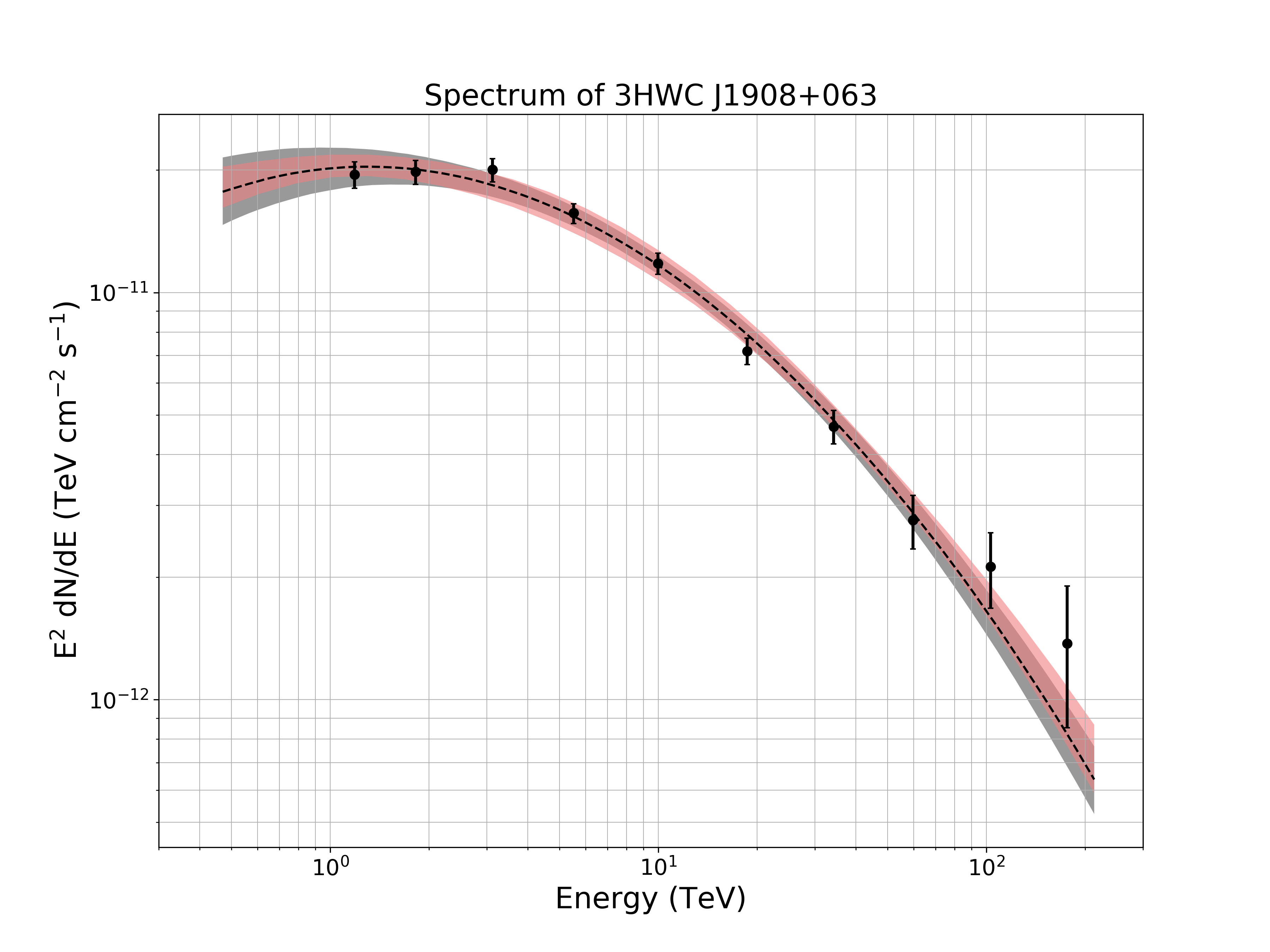}
\caption{The HAWC spectrum of 3HWC J1908+063. The grey band is the forward-folded statistical uncertainty band, while the pink band denotes systematic uncertainties related to the modeling of the detector (discussed in Section \ref{sec:systematics}). A table containing the flux points can be found in the Appendix.}
\label{fig:spectrum}
\end{figure}

Figure \ref{fig:sigAndModelMap} shows the HAWC significance map and the best-fit model for this region. Figure \ref{fig:resids} shows the residual map (the significance map of the difference between the data and the model). The best-fit parameters can be seen in Table \ref{tab:diffRes}, and the HAWC spectral energy distribution (SED) can be seen in Figure \ref{fig:spectrum}. 

A test statistic (TS) for each source in the model is computed. TS is defined as twice the likelihood ratio:
\begin{equation}
TS = 2\mathrm{ln} \left(\frac{L_{s}}{L_n} \right),
\end{equation}
where $L_{s}$ is the best-fit likelihood for the hypothesis including the source and $L_n$ denotes the null hypothesis. When Wilks' theorem~\citep{Wilks1938} is assumed, the square root of the TS can be equated to Gaussian significance.  Wilks' theorem is valid for HAWC data~\citep{2hwc}.

3HWC J1908+063 is significantly detected ($>$ 2$\sigma$) in each energy bin. Using the method presented in \cite{Crab}, this corresponds to a detection between 470 GeV and 213 TeV. The TS for this source is 1924.9. The East and West lobes of SS433 have TS values of 8.5 and 15.0, respectively. The TS values for the lobes of SS433 are lower than previously published because the analysis presented here uses an additional quality cut that removes all events whose reconstructed shower core is off the main HAWC array. This greatly increases the angular resolution of the detector, allowing for better morphological studies, but removes approximately half of the data. 

The spectral index of 3HWC J1908+063, $\alpha$, is 2.545 $\pm$ 0.026 $_{-0.04}^{+0.01}$, and the curvature parameter, $\beta$, is 0.134 $\pm$ 0.018 $_{-0.03}^{+0.01}$. The first set of uncertainties are statistical, and the second set is systematic uncertainties stemming from the mismodeling of the detector (discussed further in Section \ref{sec:systematics}).  The flux normalization, $\phi_0$ is  (1.17$\pm$0.06$\pm$0.10) $\times$ 10$^{-13}$ (TeV cm$^{2}$ s)$^{-1}$.   

The values reported here are different from those reported in the HAWC highest energy ($>$ 56 TeV) catalog~\citep{HECatalog}; most notably the source has a slighter softer spectral index than previously reported. This can be attributed to two differences in the analysis. First, in this paper we report the spectrum assuming that the reported 3HWC source location is the center of the source (right ascension = 287.05$^{\circ}$ $\pm$ 0.06$^{\circ}$, declination = 6.39$^{\circ}$ $\pm$ 0.06$^{\circ}$), while in ~\cite{catalog}, the peak of the $>$ 56 TeV emission (right ascension = 286.91$^{\circ}$ $\pm$ 0.10$^{\circ}$, declination=6.32$^{\circ}$ $\pm$ 0.09$^{\circ}$) was assumed to be the center of the source. These coordinates are 0.16$^{\circ}$ away from the 3HWC source location. Secondly, the high-energy catalog assumed a Gaussian morphology instead of the diffusion model presented here. These two models predict different fluxes, especially at the lower energy end of the spectrum.

3HWC J1908+063 has a best-fit diffusion angle, $\theta_d$, of 1.78 $\pm$ 0.08 $_{-0.02}^{+0.07}$ degrees, reported at the gamma-ray pivot energy of 10 TeV.  10 TeV gamma-rays imply $\sim$65 TeV electrons (Equation \ref{eq:electronEnergy}). Assuming that the distance to the pulsar J1907+0602, 2.37 kpc, is the same as the distance to the source, this means that the 65 TeV electrons will have diffused $\sim$74 parsecs (see Equation \ref{eq:diffRad}).

The cooling time of 65 TeV electrons is $\sim$14,000 years assuming a magnetic field of 3 $\mu$G. This is slightly smaller than the age of the source, so it is used in Equation \ref{eq:diffusion} to calculate the diffusion coefficient. The diffusion coefficient, $D$, for 65 TeV electrons is 2.92 $\times$ 10$^{28}$ cm$^{2}$/s. While this is lower than the interstellar medium (ISM) value by approximately one order of magnitude, this value is not unrealistic. Lower diffusion coefficients have been observed before ~\citep{Geminga} and a suppression of the diffusion coefficient has been predicted in certain cases, such as near TeV halos~\citep{evoli}.

\begin{figure}
\includegraphics[width=0.5\textwidth]{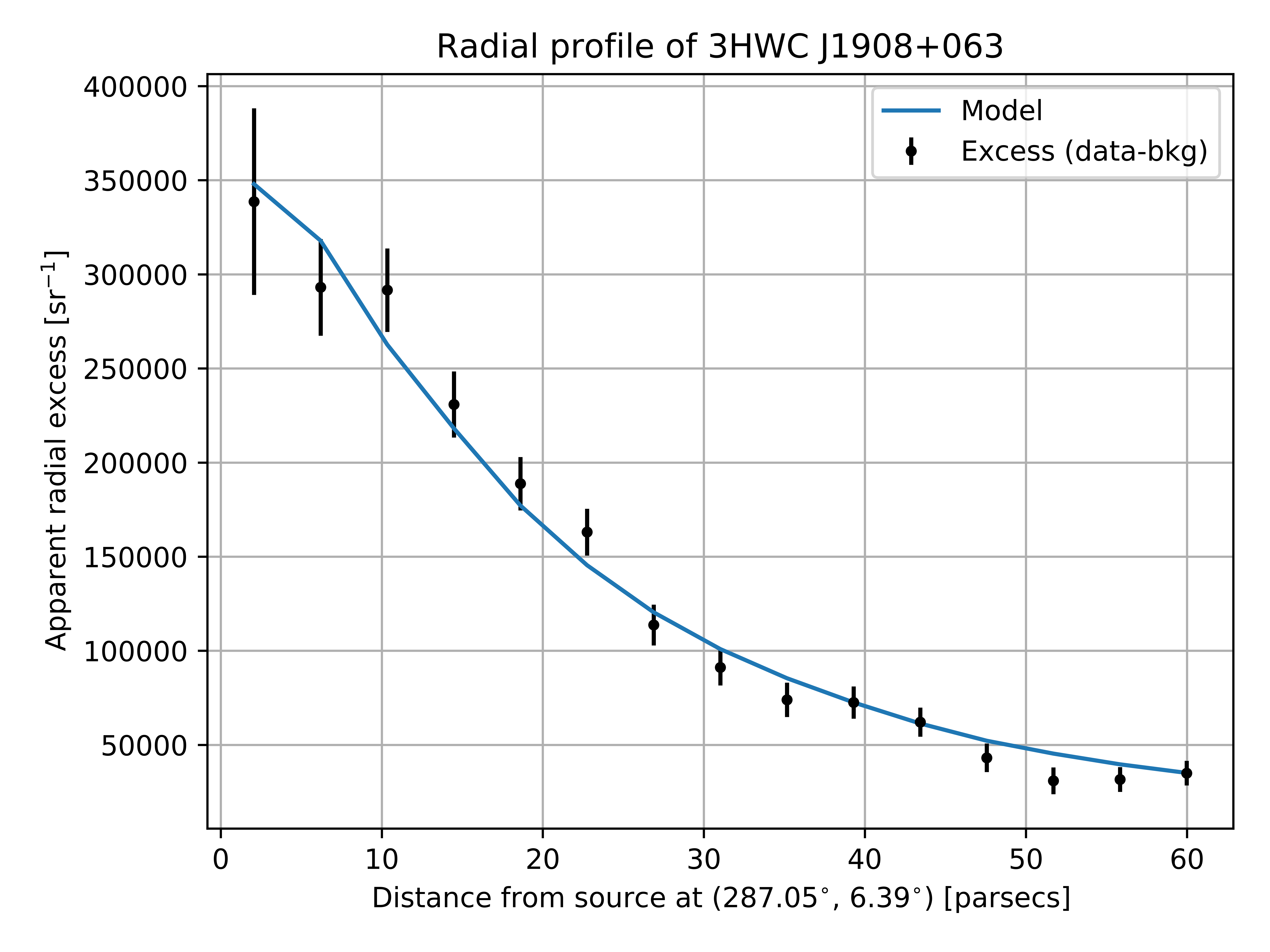}
\caption{The radial profile of 3HWC J1908+063. The blue line shows the expectation for the diffusion model. This figure assumes that the source is 2.37 kpc away, which is the distance to PSR J1907+0602.   } 
\label{fig:radial}
\end{figure}

Figure \ref{fig:radial} shows the radial profile of 3HWC J1908+063. As can be seen from the figure, the data matches the expectation for continuous injection of particles with diffusion away from the center of the source. Emission is seen out to several parsecs.

The flux normalization, $\phi_0$, for the East (West) lobe of SS433 is 2.0 $^{+1.0}_{-0.7}$ (3.0$^{+1.1}_{-0.8}$) $\times$ 10$^{-16}$ (TeV cm$^{2}$ s)$^{-1}$. These values agree with those reported in ~\cite{ss433}, within uncertainties.

A comparison between the spectrum reported here and results from other detectors is discussed in Section \ref{sec:iacts}.

\subsection{Systematic uncertainties}\label{sec:systematics}
Systematic uncertainties related to the modeling of the HAWC detector are investigated as described in ~\cite{Crab}. The effects of the detector systematic uncertainties are shown as the pink band in Figure ~\ref{fig:spectrum}. These uncertainties are a function of energy. The width of the systematic uncertainties on the flux range from $\sim _{-11\%}^{+14\%}$ at 1 TeV to $\sim _{-10\%}^{+22\%}$ at 100 TeV.

The other main source of uncertainty is the effect of Galactic diffuse emission (GDE) on the spectrum.  The base model neglects diffuse emission, as was done with past analyses of the region.  The source is located 0.8$^{\circ}$ away from the Galactic plane but is quite extended, with some emission observed at $b=0$. Here, we investigate whether diffuse emission has any effect on the reported fits. 

The GDE is modeled as a Gaussian distribution in Galactic latitude, centered on the Galactic plane. The width of the Gaussian, $\sigma$, is allowed to float in the fit.  The emission is assumed to emit according to a power-law spectrum:
\begin{equation}
\frac{dN}{dE} = \phi_0 \left(\frac{E}{\text{7 TeV}}\right)^{-\alpha}
\end{equation}

The value of $\alpha$ is fixed at 2.75, chosen to match the observed cosmic-ray spectrum around 10 TeV~\citep{diffuse,diffuseICRC}. The flux normalization, $\phi_0$, is a free parameter in the fit, along with $\sigma$.

The best-fit values for the GDE are $\phi_0$ = (1.9 $\pm$ 0.4) $\times$ 10$^{-14}$ (TeV cm$^{-2}$ s)$^{-1}$ and $\sigma$ = 0.64 $\pm$ 0.17 degrees. 

The best-fit values for 3HWC J1908+063, with diffuse emission included, are: $\theta_d$ = 1.50 $\pm$ 0.10 degrees; $\phi_0$ = (0.96 $\pm$ 0.08) $\times$ 10$^{-13}$ (TeV cm$^{-2}$ s)$^{-1}$;  $\alpha$ = 2.505 $\pm$ 0.032; and $\beta$ = 0.150 $\pm$ 0.022. 

The estimated systematic uncertainty of the diffusion angle due to diffuse emission is -15.7$\%$. The effect on the flux varies by energy, ranging from 31.2$\%$ at 1 TeV to 17.7$\%$ at 100 TeV. The effect is larger at lower energies. 

A follow-up analysis investigating the morphology of 3HWC J1908+063 is in progress. This will include an in-depth study of the energy-dependent effect of the diffuse emission. 

\subsection{Comparison to other experiments}\label{sec:iacts}

\begin{figure*}
\centering
\includegraphics[width=0.8\textwidth]{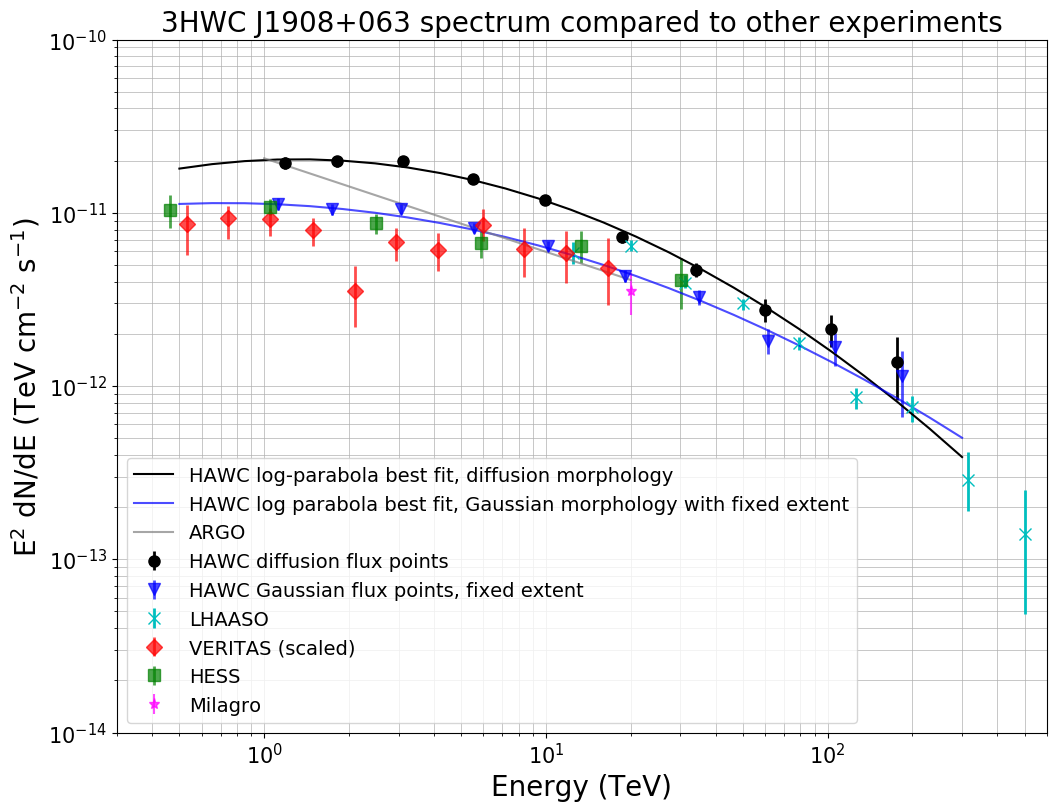}
\caption{A comparison between the spectrum reported in the main text and results from other observatories. The discrepancy between the HAWC SED and the other observatories is discussed in Section \ref{sec:iacts}. References for other experiment's measurements: VERITAS~\citep{VERITAS}, ARGO~\citep{ARGO}, Milagro~\citep{Milagro}, H.E.S.S.: Online catalog~\citep{vizier} from the H.E.S.S. Galactic plane survey~\citep{HGPS}, LHAASO~\citep{lhaaso2}. }
\label{fig:comparison2}
\end{figure*}

Figure \ref{fig:comparison2} compares the HAWC result (black line) to those obtained using other detectors, including both IACTs and other air shower arrays. HAWC measures a higher flux than IACT \replaced{experiments}{detectors}.  Differences in the field-of-view, angular resolution, and background estimation methods between IACTs and HAWC contribute to discrepancies in the measured flux~\citep{2021ApJ...917....6A}.

Additionally, as discussed in previously in ~\cite{j2019}, IACTs may extract their spectrum from a region that may be different than the measured morphology. This is a difference from HAWC, where the spectrum and morphology are fit simultaneously.

These effects combined lead to a systematic flux offset between HAWC and IACTs.  To account for this, the VERITAS points~\citep{VERITAS} have been scaled by a factor of 2.03 using the technique outlined in ~\cite{chadsthesis}.  The H.E.S.S. points, taken from the H.E.S.S. Galactic Plane survey~\citep{HGPS}, are not scaled as the H.E.S.S. collaboration took this effect into account.

It is apparent that, even with the scaling described above, the HAWC result includes more flux than the IACTs. The discrepancy is more prominent at the lower energies. 

There are also differences between HAWC and other air shower arrays. Note the Milagro result (magenta point). The ARGO spectrum~\citep{ARGO} agrees with the HAWC spectrum at the very lowest energies but quickly begins to diverge as the energy increases. It is interesting to note that the LHAASO result~\citep{lhaaso2} agrees with the HAWC result relatively well, especially at the highest energies.  

The source of this difference can be attributed to differences between the morphologies assumed by the other observatories.  HAWC is unique in using a diffusion model, which has a long tail. If the HAWC data is instead fit assuming a Gaussian morphology with the extent fixed to the IACT extent (here, 0.44 degrees is used to match the VERITAS extent), the discrepancy disappears (see the blue curve in Figure \ref{fig:comparison2}). In this study, the right ascension and declination are left fixed to the 3HWC coordinates.  Due to their background estimation methods and morphological assumptions, IACTs are not sensitive to the long, extended tails of the gamma-ray emission.  This may affect their physics conclusions. 

The HAWC Gaussian flux points are consistent with the HAWC diffusion flux points above 100 TeV. This will be important in the Modeling section (section \ref{sec:modeling}) where these data points are used to draw conclusions. 

\subsection{Potential spectral hardening at the highest energies}\label{sec:hardening}

\begin{figure}
\includegraphics[width=0.5\textwidth]{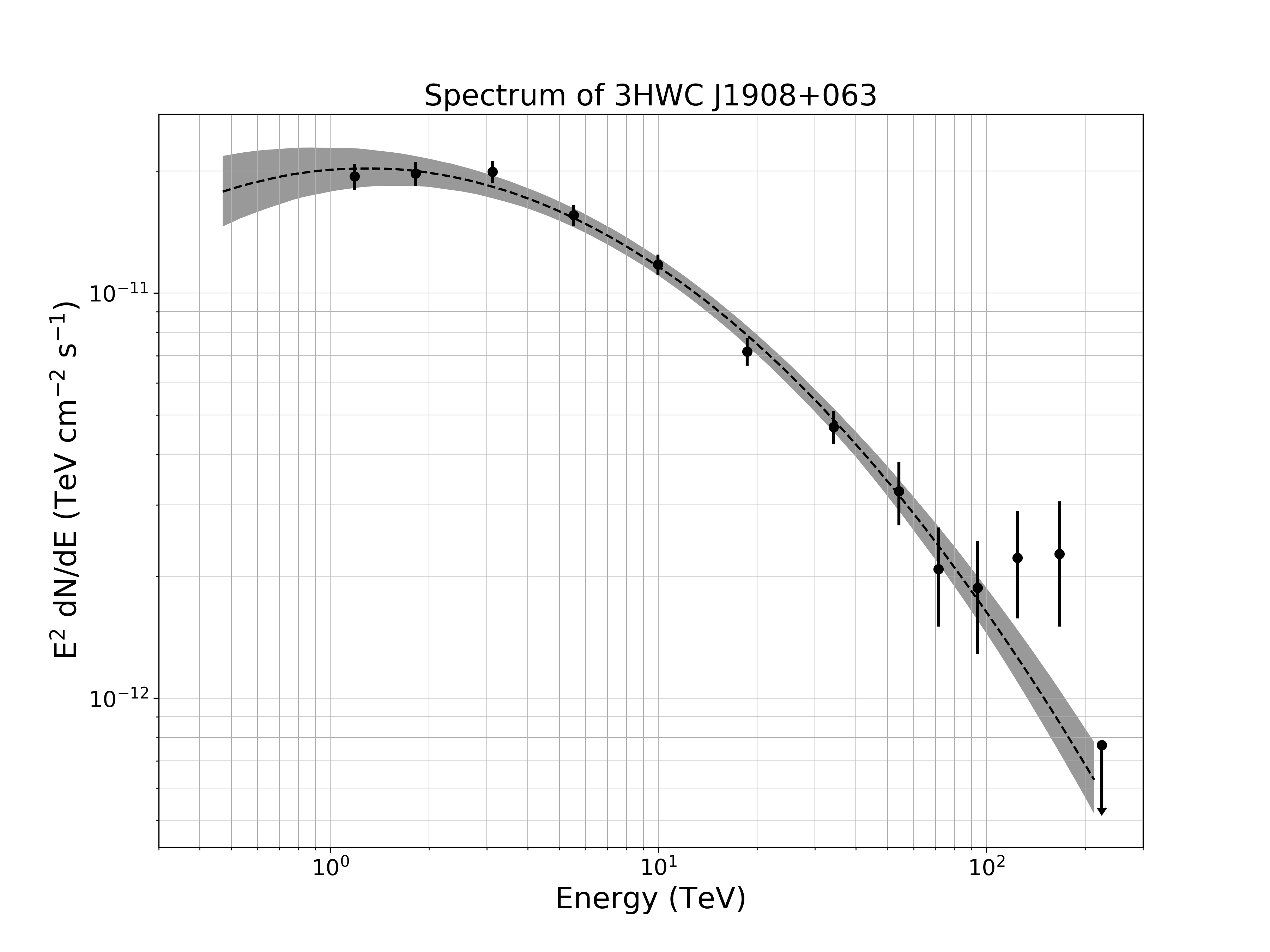}
\caption{The HAWC spectrum of 3HWC J1908+063 when the three highest-energy bins are subdivided into six sub-bins. The potential spectral hardening feature begins around 70 TeV. The source is not significantly detected in the last energy bin so a 90$\%$ confidence level upper limit is plotted. A table containing the flux points can be found in the Appendix.} 
\label{fig:spectralHardening}
\end{figure}

We look for evidence that the spectrum hardens at the highest energies. Such a discovery could indicate that there are hadronic contributions; if the source were entirely leptonic, curvature in the spectrum is expected at the highest energies due to Klein-Nishina effects~\citep{moderski}. This study is motivated by the last two flux points shown in Figure \ref{fig:spectrum}, which appear as if they are beginning to deviate from the best-fit log parabola shape by exhibiting a slight flux enhancement over expectation. It is advantageous to have more flux points at the highest energies because this is where the deviations between leptonic and hadronic models will have the most prominent differences. To investigate this, the three highest-energy bins, which each span a quarter-decade in log$_{10}$-energy space ($\hat{E}>$56 TeV) are subdivided into six smaller bins, each spanning an eighth of a decade in width, before rerunning the spectral fit.

Figure \ref{fig:spectralHardening} shows the spectrum of 3HWC J1908+063 with the highest energy bins split into sub-bins. As in the nominal analysis, a log-parabola spectral shape and the diffusion morphology are used. The overall forward-folded fit results are compatible with those presented in Section \ref{sec:bestfitresults}: $\theta_d$ = 1.77 $\pm$ 0.08; $\phi_0$ = 1.16 $\times$ 10$^{-13}$; $\alpha$ = -2.544 $\pm$ 0.026 and $\beta$ = 0.133 $\pm$ 0.018. All uncertainties are statistical only.  

A potential spectral hardening feature is visible in the flux points by eye, beginning around $\sim$ 70 TeV. The source is not significantly detected in the last energy bin (234 $< \hat{E}$ $<$ 316 TeV; TS = 0.39) so a 90$\%$ confidence level upper limit is computed. In all of the other sub-bins, the TS is $>$ 16 ($>$ 4$\sigma$).

This spectral hardening feature is presently not significant. The last two flux points in Figure \ref{fig:spectralHardening} that are significantly detected are only 1.5$\sigma$ and 1.8$\sigma$ away from the best-fit log-parabola line. Adding these two points in quadrature, the total significance of this potential feature is 2.3$\sigma$. Improvements to the HAWC reconstruction algorithms or studies using future detectors will decrease the uncertainties and make a more definitive statement. It is worth noting that this feature is remains visible by eye if detector response functions designed to probe systematic uncertainties related to the mis-modeling of the HAWC detector are used, so it is unlikely to be an instrumental effect 

If this feature is shown to be significant in the future, it would be strong evidence for hadronic processes associated with this source. See the Modeling section (Section \ref{sec:modeling}) for a further discussion of this.

If the $>$ 100 TeV gamma rays originate from hadronic processes, this requires proton energies corresponding to the knee of the cosmic-ray spectrum. It is unclear what the origin of these cosmic rays would be. One scenario is that they could be from the SNR interacting with molecular clouds in the region. Another scenario is that hadronic acceleration mechanisms occur in PWN. While this does not occur in conventional models, this possibility has been explored in the literature~\citep{Bednarek,dipalma}.

\section{Theoretical Modeling}\label{sec:modeling}

Here, we model the TeV data to explore possible emission mechanisms. In general, the lack of available multi-wavelength data means that the parameter space is largely unconstrained. For example, the magnetic field in the region is unknown.

Different conclusions can be reached depending on what modeling assumptions are made. We begin by showing that a one-population hadronic model is unlikely, although it cannot be completely ruled out. We then present two models: one where the radiation energy density dominates over the magnetic field energy density ($u_{rad} \gg u_{B}$), and one where the converse is true ($u_{B} \gg u_{rad}$).

Note that throughout this section, the phrases ``one-population" and ``two-population" refer to the number of particle populations present in the TeV range.  As will be discussed later, there is an additional component in the GeV range, based on recent Fermi-LAT data~\citep{fermiPWN}.

\subsection{One-population hadronic model}\label{sec:hadrons}

A purely hadronic model is disfavored due to a lack of sufficient energy to power the hadronic emission. To show this, first we calculate the target proton density in the region. Molecular clouds in the region can affect this density. There are several known clouds near 3HWC J1908+063. Here, we use data from CO survey published in ~\cite{DAME}. Data from the 35 to 50 km/s range is used, as this corresponds to the distance to PSR J1907+0602. Atomic gas is also considered, using the HI4PI survey~\citep{HI4PI}. Assuming a circle of radius 1$^{\circ}$ centered on the HAWC source, the average combined column density from these two surveys totals 9.9 $\times$ 10$^{21}$ cm$^{-2}$. A rough estimate of the mass can then be obtained. The radius of the 1$^{\circ}$ circle is 41.36 pc, assuming that the gas is 2.37 kpc away (the distance to PSR J1907+0602).  Then, 
\begin{equation}\label{eq:mass}
m = \pi r^2 N m_p,
\end{equation}
where $m$ is the total mass in the region, $r$ is the radius, $N$ is the column density and $m_p$ is the mass of the proton.  The total mass is 8.5 $\times$ 10$^{35}$ kg, which leads us to a proton density ($n$) of $\sim$60 protons cm$^{-3}$. This is considerably higher than the typical ISM value of 1 cm$^{-3}$. This calculation assumes that the gas is spherical. 

Assuming that the source is purely hadronic and that the population of protons is trapped within the cloud, the total energy in nonthermal hadrons is 
\begin{equation}
W_{pp} = L_{\gamma} \frac{t_{pp}}{\eta_n}
\end{equation}
where $L_{\gamma}$ is the gamma ray luminosity:
\begin{equation}\label{eq:lum}
L_{\gamma} = 4 \pi d^2 \int_{E_{min}}^{E_{max}} \frac{dN}{dE} E dE,
\end{equation}
$t_{pp}$, the proton energy loss timescale, is equal to 2$\times$10$^{15}n^{-1}$ sec cm$^{-3}$ and $\eta_n$ is equal to 1.5.  The parameter $\eta_n$ accounts for the fact that there are nuclei heavier than hydrogen in both cosmic rays and interstellar matter~\citep{pevatron}. In Equation \ref{eq:lum}, $\frac{dN}{dE}$ is the gamma-ray flux per energy. It is not related to $N$ (the column density) from Equation \ref{eq:mass}.

From the fitted HAWC spectrum (see Section \ref{sec:bestfitresults}), $E_{min}$ is 470 GeV and $E_{max}$ is 213 TeV.  Using the best-fit values for the spectrum to calculate $L_{\gamma}$ and 60 protons cm$^{-3}$ for the number density, $n$, we obtain a value of 1.7 $\times$ 10$^{48}$ erg for $W_{pp}$ over the energy range of HAWC. This is the total energy available based on the gas in the region. However, in order to explain the HAWC data points, a simple hadronic model requires a total energy on the order of $\sim$10$^{50}$ erg, assuming that $n$ is $\sim$60 cm$^{-3}$ and the parent proton spectrum can be modeled by a single power law. This is two orders of magnitude higher in energy than the available energy calculated above, so the hadronic model is difficult to explain. 

We cannot completely reject a one-population model because the lower energies are unconstrained by the HAWC data. The energy requirement of 10$^{50}$ ergs assumes that the protons can be modeled by a power law all the way down to the lowest energies. However, if the spectrum of the protons is instead curved, it is conceivable that the total energy required could be lower, as calculating the energy budget involves integrating over the spectrum.

\subsection{One-population Inverse Compton model}\label{sec:ic}

An inverse Compton (IC) origin of the emission is attractive given the absence of a clear correlation with target material.

In a single zone/population model and in the case that synchrotron losses are dominant at all energies, the equilibrium high-energy electron spectrum is steep and Klein-Nishina (KN) suppression results in a very steep IC spectrum which is not consistent with the HAWC data. Alternatives to this picture are the existence of more than one emission component (see Section \ref{sec:density}) or that the magnetic field energy density lies below that of radiation fields such that IC cooling effects become important. This case was considered in detail by \cite{breuhaus, Breuhaus_et_al_2021_LHAASO} and here we adopt the same approach.

Electrons following an exponential cutoff power-law
\begin{equation}
\frac{\text dN}{\text dE} = N_0 \left(\frac{E}{1 \mathrm{erg}}\right)^{-\alpha} \exp\left(-\frac{E}{E_{\text{cut}}}\right)    
\end{equation}
are injected into a region with constant, isotropic and homogeneous radiation fields and a fixed magnetic field of 3 $\mu$G. The radiation field adopted is that of the large scale Galactic radiation field model of \cite{Popescu2017} at the location of the possibly associated pulsar PSR\,1907+0602 (plus the CMB). The injection at a constant rate takes place for 19.5 kyr, the characteristic age of the pulsar. A fit to the data with $N_0$, the injection index $\alpha$ and the cutoff energy $E_{\text{cut}}$ as free parameters gives $N_0 = (1.3 \pm 0.2) \times 10^{36}$ ergs/second and $\alpha$ = $2.68\pm 0.04$. The cutoff energy $E_{\text{cut}}$ can only be constrained to be larger than 610 TeV with 95$\%$ confidence and therefore a value of $E_{\text{cut}} = $10 PeV is used. The quoted values correspond to a mean electron/positron luminosity of 50$\%$ of the current spin-down power of the pulsar injected above 1 TeV. For the numerical calculations we made use of the open source Code GAMERA \citep{gamera}.

\begin{figure}
    \centering
    \includegraphics[width=0.5\textwidth]{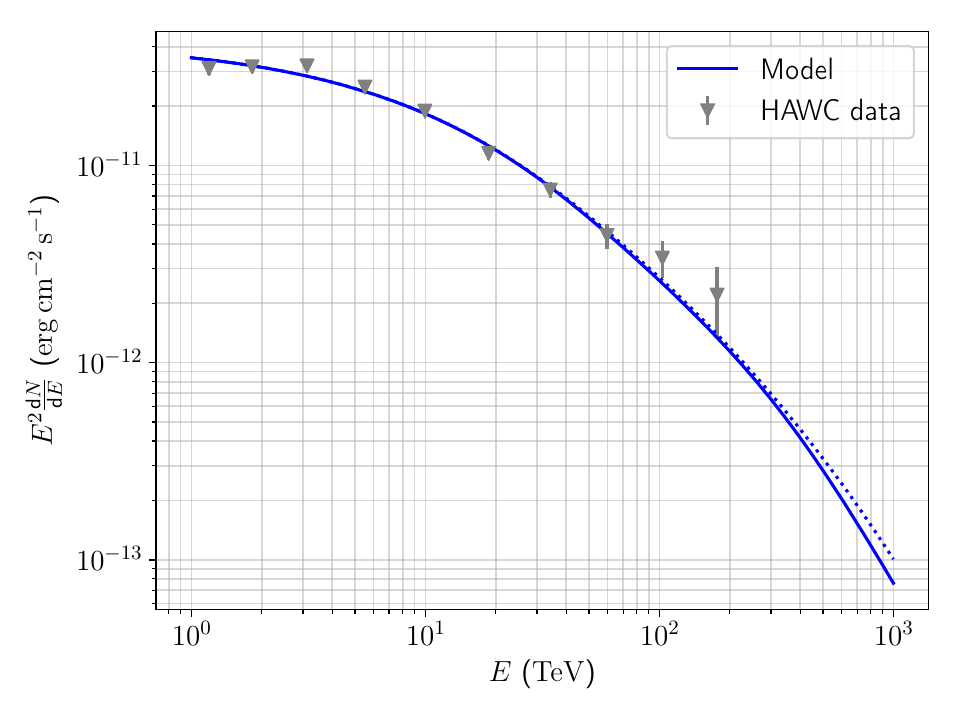}
    \caption{Inverse Compton model fit to the HAWC spectra data for J1908. The solid/dashed lines show the spectrum with/without the effects of $\gamma$-$\gamma$ absorption on the CMB and Galactic radiation fields. }
    \label{fig:model}
\end{figure}

To account for absorption of the $\gamma$-ray emission due to pair production in the interstellar photon fields on the way to Earth, the same radiation model together with the CMB was used. The model is shown in Figure \ref{fig:model}.

\subsection{Two-population models}\label{sec:density}

Assume instead that there is a region near the PWN where the magnetic field is high ($>$10 $\mu$G) and there is a TeV halo driven by the pulsar, so that $u_{B}$ is very high. Electrons have been continuously injected with constant power over the lifetime of the system. This leads to a complete dominance of synchrotron losses over Inverse Compton losses. Under these conditions, one population of leptonic particles cannot explain the TeV emission. The inverse Compton component suffers from considerable Klein-Nishina effects at the highest energies, which suppresses the high-energy gamma-ray flux. 

In this section, we explore the possibility that there are two populations responsible for the TeV particles present in the emission region: a primary population of non-thermal electrons that is responsible for the bulk of the emission, and a secondary nonthermal particle population that dominates the emission above 50 TeV. The secondary component may be either of leptonic or hadronic origin. 

Due to the unconstrained parameter space, the number of free parameters in the models described in this section are larger than the number of data points. Therefore, both models presented here are merely possible combinations of parameters that describe the data well. Both models assume that the source is located 2.37 kpc from the Earth and the ISM magnetic field strength is assumed to be 3 $\mu$G. Only contributions from the CMB photon field are considered. The spin-down power of the pulsar is assumed to be 2.8 $\times$ 10$^{36}$ ergs/s. 

In both two-population models presented here, the first component is a nonthermal electron distribution with a broken-power law shape.  The electron spectrum is given by:
\begin{equation}\label{eq:onepop}
\frac{dn_{e}}{d\gamma_e} = \begin{cases}
n_{e0} \left(\frac{\gamma_e}{\gamma_{e0,b}}\right)^{-p_{e0}} e^{-\gamma_e/\gamma_{e0,cut}} & \gamma_e < \gamma_{e0,b}\\ 
n_{e0} \left(\frac{\gamma_e}{\gamma_{e0,b}}\right)^{-(p_{e0}+1)} e^{-\gamma_e/\gamma_{e0,cut}} & \gamma_e > \gamma_{e0,b}
\end{cases},
\end{equation}
where $\gamma_e$ is the electron Lorentz factor, $n_{e0}$ is the spectral normalization factor, $\gamma_{e0,b}$ is the Lorentz factor at the cooling break, $\gamma_{e0,cut}$ is the electron Lorentz factor at the spectral cutoff, and $p_{e0}$ is the spectral index. The spectral index changes by one unit after the break, as is expected from radiative cooling.  The spectrum begins at  $\gamma_{e0,min}$ = 10$^3$ ($\sim$ 500 Mev). 

The nonthermal electron energy density in the emission region is given by
\begin{equation}
u_{e,nth}=\int_1^{\infty}\frac{dn_e}{d\gamma_e} \gamma_e m_ec^2 d\gamma_e~~.
\end{equation}
Values of $\gamma_{e0,b}=5\times 10^7$ (26 TeV), $\gamma_{e0,cut}=1.2\times10^8$ (61 TeV), and $p_{e0}=2$ provide a good description of the data. Assuming that the emission region is homogeneous, the total nonthermal electron energy is $1.6\times 10^{48}~\rm{erg}$. The location of the cooling break ($\gamma_{e0,b}$) corresponds to a cooling time of $\tau_{cool}=2.57\times10^{4}$ years, which is roughly the same as the characteristic age of the pulsar.  93$\%$ of the spin-down power of the pulsar is contained in this component.

\subsubsection{Two-population purely leptonic model}

To describe the hard spectrum at $\gtrsim 50~\rm{TeV}$, we introduce a second nonthermal electron population. The second population is assumed to be from a more recent active phase of the source and has the form:
\begin{equation}
\frac{dn_{e}}{d\gamma_e} = n_{e1} \gamma_e^{-p_{e1}} e^{-\gamma_{e}/\gamma_{e1,cut}}~~,
\end{equation}
where the spectrum extends from $\gamma_{e1,min}=10^3$ ($\sim$500 MeV) to $\gamma_{e1,cut}= 10^{9}$ ($\sim$500 TeV) and $p_{e1}=2$. The total energy in this component is 4 $\times$ 10$^{46}$ erg. This is only a fraction of the primary leptonic component.  We assume that the electrons in this second component were injected in the last $\sim$2000 years. This component contains 2$\%$ of the spin-down power of the pulsar. 

Note that the second pulsar in the region, PSR J1907+0631, is unlikely to be able to accelerate particles to the energies discussed here, so this second component is still likely associated with PSR J1907+0602. 

This model is shown as the thick blue line in Figure \ref{fig:comparison}. Note that this model violates the X-ray upper limit from XMM-Newton. However, we note that the XMM-Newton upper limit is extracted from a region that is smaller (a 0.75$^{\circ}$ by 0.75$^{\circ}$ square) than the HAWC extent. The observation was centered on the pulsar, which is $\sim$0.3$^{\circ}$ away from the centroid of the HAWC source. It is difficult for X-ray satellites to observe regions that are a degree across, as this source is, due to their relatively small fields-of-view. Nevertheless, the second electron population results in extra flux in the X-ray band, which may be examined by future X-ray telescopes.

\subsubsection{Two-population hybrid lepto-hadronic model}\label{sec:leptohadronic} 

\begin{figure*}
\centering
\includegraphics[width=0.8\textwidth]{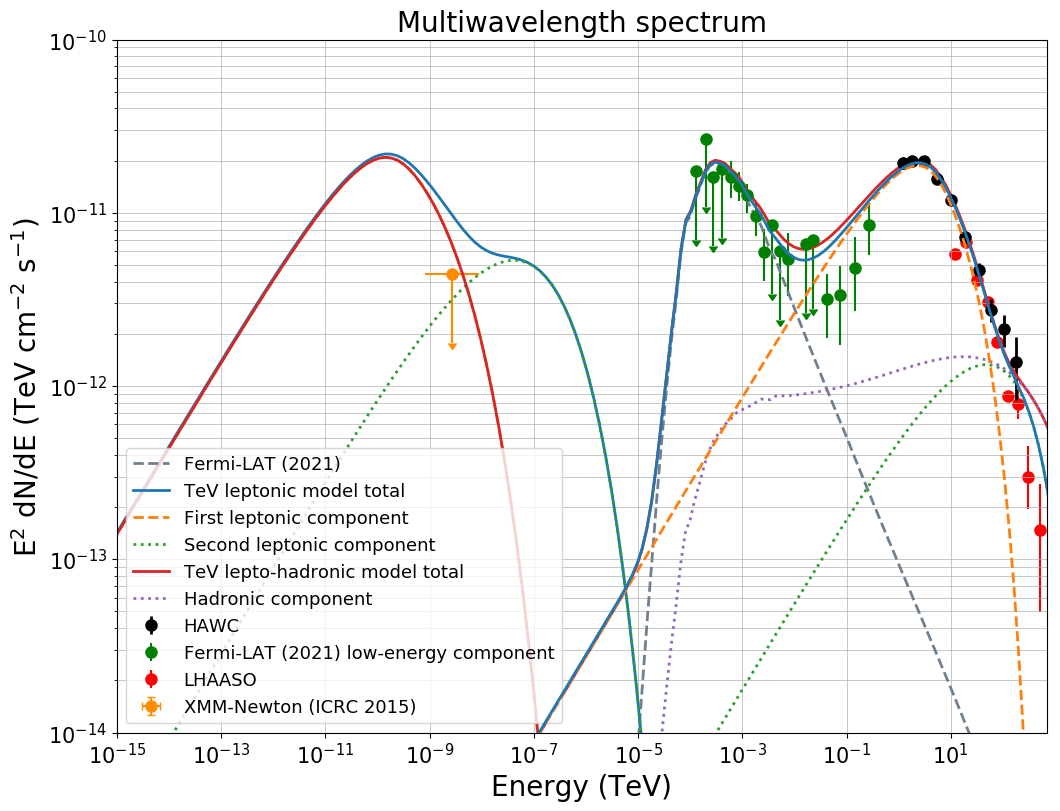}
\caption{The best-fit multi-population TeV leptonic (the thick red line, which is the sum of the dashed orange, dotted green lines, and dashed grey lines) and the TeV lepto-hadronic models (thick blue line, which is the sum of the dashed orange, dotted purple, and dashed grey lines). Both models contain an additional component in the GeV range, as recently discovered using Fermi-LAT data (see Section ~\ref{sec:fermiCompare} for a discussion). The LHAASO points, from ~\cite{lhaaso2}, are shown for comparison purposes only. Note that one-population TeV models can also fit the data well (see the disucssion in Section \ref{sec:ic}).}
\label{fig:comparison}
\end{figure*}

A hybrid lepto-hadronic model can also explain the HAWC data well. In this scenario, the extra nonthermal particle distribution consists of protons. The TeV gamma-ray emission is created via proton-proton collision. The hadronic population is modeled as follows: 
\begin{equation}
\frac{dn_p}{d\gamma_p} = n_{p} \gamma_p^{-{p_p}} e^{-\gamma_p /\gamma_{p,cut}}
\end{equation}
where the proton spectrum extends from $\gamma_{p,min}=1$ ($\sim$900 MeV) to $\gamma_{p,cut}=10^7$ ($\sim$ 9 PeV), and the spectral index $p_p$ is 2. Assuming these model parameters, the energy in protons is 1.5 $\times$ 10$^{48}$ ergs. Note that this is close to the energy in the electrons. The solid red line in Figure \ref{fig:comparison} shows the predicted flux for the lepto-hadronic model. There is much less tension with the X-ray upper limit than in the two-population leptonic model. This is expected, because the pair synchrotron emission, a byproduct of proton-proton collision, gives trivial contribution to the emission due to the very low ISM magnetic field (3 $\mu G$).  Note that this model also overshoots the X-ray upper limit, although only slightly. 

A major issue is how to accelerate nonthermal protons in the source. There are two possible scenarios. The first scenario is that the protons were accelerated when the supernova exploded, and then diffused out of the shock. The other scenario is that the protons were accelerated at the termination shock by the pulsar wind. In either scenario, the nearby molecular clouds provide a target for the gamma-ray emission. 

\subsubsection{Comparison of two-population model to recent Fermi-LAT results}\label{sec:fermiCompare}

Extended GeV emission has recently been detected using data from the Fermi-LAT~\citep{fermiPWN}. This GeV spectrum consists of two components: a softer component below 10 GeV which is likely associated with molecular clouds near SNR G40.5-0.5, and a harder, higher-energy component that is likely inverse Compton in origin and associated with the PWN.

The modeling presented in the preceding two sections was developed using TeV data only. In Figure ~\ref{fig:comparison}, we show the sum of each of the two-population TeV models with the lower-energy GeV component. Since HAWC is not sensitive to multi-GeV energies, we simply use the parameters from ~\cite{fermiPWN}. The second, higher-energy GeV component is not shown since it is simply the extrapolation down to lower energies of the inverse Compton TeV component. 

The higher-energy Fermi-LAT data points do not smoothly connect to the HAWC data points or modeling. However, they do smoothly connect with IACT measurements from H.E.S.S. and VERITAS. ~\cite{fermiPWN} hypothesize that this is due to background estimation differences between Fermi-LAT/IACTs and HAWC, a statement consistent with the discussion in Section \ref{sec:iacts}.

The addition of the lower-energy Fermi-LAT component indicates that there may be as many as three particle populations across all wavelengths: the GeV SNR/molecular cloud emission, the GeV/TeV inverse Compton emission, and then the third component that is prominent above 56 TeV. This third component, if hadronic, may not be linked to the SNR. Instead, it could originate from more exotic mechanisms such as hadron acceleration in PWN, as has been proposed by ~\cite{hadron1}, ~\cite{Bednarek}, and ~\cite{dipalma}, among others. This component could also result from interactions between the PWN and the SNR.  

\subsubsection{Comparing the different two-population models}\label{sec:compareModels}

Figure \ref{fig:comparison} shows a comparison of the two different multi-population models.

In the event that there are two populations of particles responsible for the TeV emission, the current energy resolution of HAWC prevents us from definitively saying whether the two-population leptonic model is preferred over the lepto-hadronic model. Both models fit the data above 50 TeV within the uncertainties on the flux points. This strengthens the case for future detectors, both proposed and currently under construction, that will have greater sensitivity above 100 TeV. This is discussed further in Section \ref{sec:multi}.

\section{Implications for multi-wavelength and multi-messenger experiments}\label{sec:multi}

The modeling presented above has implications for the detection of this source by multi-wavelength and multi-messenger detectors.

The LHAASO experiment~\citep{lhaaso} is able to probe higher energy ranges than HAWC. A recent publication by the collaboration detected ultra-high-energy emission from the region; the maximum photon energy detected was 440 TeV~\citep{lhaaso2}. An in-depth study of the spectrum and morphology at the highest energies using LHAASO data could help distinguish between the models presented in this work. Note that the Inverse Compton model and the two-population models predict different fluxes at the very highest energies. Neither of our models are able to distinguish between the highest-energy emission mechanisms, as the uncertainties on the highest-energy flux points are presently too large.

Additionally, the two-population leptonic model and the lepto-hadronic model predict very different amounts of synchrotron emission in the keV to MeV energy bands. Proposed detectors such as AMEGO~\citep{kierans} will be important in distinguishing emission mechanisms.  Additional X-ray to GeV gamma-ray observations will allow us to pinpoint the spectral shape and energy budget.  

Additional x-ray observations may also allow us to determine the magnetic field in the region, allowing us to differentiate between one-population and two-population models. For example, if it is shown that there is considerable synchrotron emission in the region, the magnetic field is likely high (tens of $\mu$G) and two particle distributions are required to explain the spectrum. 

Most hypotheses of whether the IceCube Neutrino Observatory will see this source assume that the emission is entirely hadronic~\citep{gonzalezgarcia,Halzen2017}. As discussed in Section \ref{sec:hadrons}, a pure hadronic scenario seems unlikely based on the energy budget. Here, we estimate whether IceCube will see 3HWC J1908+063 in the lepto-hadronic model, where the hadronic component accounts for approximately 10$\%$ of the TeV flux and can be approximated as a power-law. In the case of either a one-population leptonic or two-population leptonic model, no neutrinos are expected.

Equations 2 and 3 of \cite{Halzen2017} show the relation between a gamma-ray flux and the corresponding neutrino flux.  If the TeV gamma-ray flux due to hadrons is

\begin{equation}\label{eq:gammaflux}
\frac{dN_{\gamma}(E_{\gamma})}{dE_{\gamma}} = k_{\gamma} \left(\frac{E_{\gamma}}{\mathrm{1 TeV}}\right)^{-\alpha} e^{-\sqrt{\frac{E_{\gamma}}{E_{cut,\gamma}}}}
\end{equation}
where $E_{\gamma}$, $k_{\gamma}$, $\alpha_{\gamma}$, and $E_{cut,\gamma}$ are the gamma-ray energy, flux normalization, spectral index, and energy cutoff respectively, then the corresponding neutrino flux is 

\begin{equation}\label{eq:neutrinoflux}
\frac{dN_{\nu_{\mu} + \bar{\nu_{\mu}}(E_{\nu})}}{dE_{\nu}} = k_{\nu} \left(\frac{E_{\nu}}{\mathrm{1 TeV}}\right)^{-\alpha_{\nu}} e^{-\sqrt{\frac{E_{\nu}}{E_{cut,\nu}}}}
\end{equation}
Here, $k_{\nu}$ = (0.694 - 0.16$\alpha_{\gamma}$)$k_{\gamma}$, $\alpha_{\nu}$ = $\alpha_{\gamma}$ and $E_{cut,\nu}$ = 0.59$E_{cut,\gamma}$.

Since the leptonic contribution to the gamma-ray flux is not expected to contribute to the neutrino flux, we use only the hadronic contribution when computing the gamma-ray flux in Equation \ref{eq:gammaflux}. This component has approximately the following values: $\alpha_{\gamma}$ = 2, $k_{\gamma}$ = 2 $\times$ 10$^{-12}$ (TeV cm$^2$ s)$^{-1}$, and no gamma-ray cutoff ($E_{cut,\gamma}$ = $\infty$).  The lack of cutoff allows us to neglect the exponential terms in Equations \ref{eq:gammaflux} and \ref{eq:neutrinoflux}.

Using the conversion between $k_{\nu}$ and $k_{\gamma}$ given above, $k_{\nu}$ is equal to $\sim$7.5 $\times$ 10$^{-13}$ (TeV cm$^2$ s)$^{-1}$.

Now that the expected neutrino flux has been computed, we can discuss whether IceCube will see this source. We assume that the neutrino source, if it exists, is extended. IceCube's discovery potential for extended sources is given in Figure 3 of ~\cite{IceCubeExtended}. The discovery potential increases as the size of the source decreases, but even if the source is only 1$^{\circ}$ across, the predicted neutrino flux is approximately an order of magnitude below the discovery potential. 

The proposed next-generation IceCube-Gen2 will have a better discovery potential and may be able to detect this source if the lepto-hadronic hypothesis is true. The predicted neutrino flux is near the discovery potential, as can be seen in Figure 8 of ~\cite{gen2}. In the absence of a detection in ten years of IceCube-Gen2 observations, it will be possible to place constraints on the hadronic emission. 

\section{Conclusions}\label{sec:conclusions} 

We report HAWC observations of the spectrum of the ultra-high-energy source 3HWC J1908+063, which emits to at least 200 TeV. The source is modeled using an electron diffusion model. There is potential spectral hardening observed at the highest energies, although more data is needed to test this hypothesis.

We investigate the origins of the TeV gamma-ray emission and conclude that an entirely hadronic scenario is unlikely. Due to the unconstrained parameter space, one-population leptonic, two-population leptonic, and lepto-hadronic models are all allowed. In the case of a lepto-hadronic model, the hadronic contribution is most important at the highest energies. Multi-messenger and multi-wavelength observations will be important in distinguishing between these two scenarios. 

\acknowledgments

We acknowledge the support from: the US National Science Foundation (NSF); the US Department of Energy Office of High-Energy Physics; the Laboratory Directed Research and Development (LDRD) program of Los Alamos National Laboratory; Consejo Nacional de Ciencia y Tecnolog\'ia (CONACyT), M\'exico, grants 271051, 232656, 260378, 179588, 254964, 258865, 243290, 132197, A1-S-46288, A1-S-22784, c\'atedras 873, 1563, 341, 323, Red HAWC, M\'exico; DGAPA-UNAM grants IG101320, IN111716-3, IN111419, IA102019, IN110621, IN110521; VIEP-BUAP; PIFI 2012, 2013, PROFOCIE 2014, 2015; the University of Wisconsin Alumni Research Foundation; the Institute of Geophysics, Planetary Physics, and Signatures at Los Alamos National Laboratory; Polish Science Centre grant, DEC-2017/27/B/ST9/02272; Coordinaci\'on de la Investigaci\'on Cient\'ifica de la Universidad Michoacana; Royal Society - Newton Advanced Fellowship 180385; Generalitat Valenciana, grant CIDEGENT/2018/034; Chulalongkorn University’s CUniverse (CUAASC) grant; Coordinaci\'on General Acad\'emica e Innovaci\'on (CGAI-UdeG), PRODEP-SEP UDG-CA-499; Institute of Cosmic Ray Research (ICRR), University of Tokyo, H.F. acknowledges support by NASA under award number 80GSFC21M0002. We also acknowledge the significant contributions over many years of Stefan Westerhoff, Gaurang Yodh and Arnulfo Zepeda Dominguez, all deceased members of the HAWC collaboration. Thanks to Scott Delay, Luciano D\'iaz and Eduardo Murrieta for technical support.

\newpage
\appendix

\section{HAWC data points}
The two tables in this section (Table \ref{table:hawcDataPoints} and Table \ref{table:hawcDataPoints2}) contain the HAWC flux points for 3HWC J1908+063. 

\begin{table}
\begin{center}
\caption{HAWC flux points for the nominal fit}
\begin{tabular}{|c | c |c|}
\hline
Energy & $E^2$ flux (TeV cm$^{-2}$ s$^{-1}$) & Test statistic \\ 
\hline\hline
1.19 & 1.95$^{+0.14}_{-0.15}$ $\times 10^{-11}$ & 187.88\\
1.82 & 1.98$^{+0.14}_{-0.13}$ $\times$ $10^{-11}$ & 216.88 \\
3.12 & 2.00$\pm 0.13$ $\times$ $10^{-11}$ & 265.03 \\
5.52 & 1.57$\pm 0.09$ $\times$ $10^{-11}$ & 351.59 \\ 
9.96 & 1.18$\pm 0.07$ $\times$ $10^{-11}$ & 372.40 \\ 
18.65 & 7.19$^{+0.55}_{-0.53}$ $\times$ $10^{-12}$ & 234.39 \\ 
34.17 & 4.70$^{+0.46}_{-0.45}$  $\times$ $10^{-12}$ & 178.21 \\
59.71 & 2.75$^{+0.43}_{-0.42}$ $\times$ $10^{-12}$ & 74.18 \\ 
103.07 & 2.13$^{+0.44}_{-0.47}$  $\times$ $10^{-12}$ & 42.98 \\ 
176.38 & 1.38$\pm 0.54$ $\times$ $10^{-12}$ & 13.37 \\ 
\hline
\end{tabular}
\\
This table contains the HAWC flux points for the nominal best fit, which is shown in Figure \ref{fig:spectrum}.
\label{table:hawcDataPoints}
 \end{center}
\end{table}

\begin{table}
\begin{center}
\caption{HAWC flux points for the sub-divided fit}
\begin{tabular}{|c | c |c|}
\hline
Energy & $E^2$ flux (TeV cm$^{-2}$ s$^{-1}$) & Test statistic \\ 
\hline\hline
1.19 & 1.94$\pm 0.15$ $\times 10^{-11}$ & 187.82\\
1.82 & 1.98$^\pm 0.14$ $\times$ $10^{-11}$ & 216.59 \\
3.12 & 1.99$^{0.13}_{-0.12}$ $\times$ $10^{-11}$ & 265.04 \\
5.52 & 1.56$\pm 0.09$ $\times$ $10^{-11}$ & 351.51 \\ 
9.96 & 1.18$\pm 0.07$ $\times$ $10^{-11}$ & 372.20 \\ 
18.65 & 7.17$\pm 0.54$ $10^{-12}$ & 234.28 \\ 
34.18 & 4.66$^{+0.45}_{-0.43}$ $\times$ $10^{-12}$ & 178.19 \\
54.08 & 3.24$\pm 0.60$ $\times$ $10^{-12}$ & 53.24 \\ 
71.30 & 2.07$^{+0.58}_{-0.56}$ $\times$ $10^{-12}$ & 22.29 \\ 
93.89 & 1.88$^{+0.57}_{-0.56}$ $\times$ $10^{-12}$ & 19.12 \\
124.36 & 2.23$^{+0.66}_{-0.65}$ $\times$ $10^{-12}$ & 23.80 \\
166.97 & 2.26$^{+0.80}_{-0.79}$ $\times$ $10^{-12}$ & 19.30 \\
224.45 & 7.91 $\times 10^{-13}$ & 0.39 \\
\hline
\end{tabular}
\\
This table contains the HAWC flux points for the scenario where the highest-energy bins are subdivided into smaller energy bins to look for evidence of spectral hardening, which is shown in Figure \ref{fig:spectralHardening}.  The last point is an upper limit.
\label{table:hawcDataPoints2}
 \end{center}
\end{table}

\bibliography{bib}

\begin{thebibliography}{}
\expandafter\ifx\csname natexlab\endcsname\relax\def\natexlab#1{#1}\fi
\providecommand{\url}[1]{\href{#1}{#1}}
\providecommand{\dodoi}[1]{doi:~\href{http://doi.org/#1}{\nolinkurl{#1}}}
\providecommand{\doeprint}[1]{\href{http://ascl.net/#1}{\nolinkurl{http://ascl.net/#1}}}
\providecommand{\doarXiv}[1]{\href{https://arxiv.org/abs/#1}{\nolinkurl{https://arxiv.org/abs/#1}}}

\bibitem[{{Aartsen} {et~al.}(2019){Aartsen}, {Ackermann}, {Adams}, {Aguilar},
  {Ahlers}, {Ahrens}, {Altmann}, {Andeen}, {Anderson}, {Ansseau}, {Anton},
  {Arg{\"u}elles}, {Auffenberg}, {Axani}, {Backes}, {Bagherpour}, {Bai},
  {Barbano}, {Barron}, {Barwick}, {Baum}, {Bay}, {Beatty}, {Becker Tjus},
  {Becker}, {BenZvi}, {Berley}, {Bernardini}, {Besson}, {Binder}, {Bindig},
  {Blaufuss}, {Blot}, {Bohm}, {B{\"o}rner}, {Bos}, {B{\"o}ser}, {Botner},
  {Bourbeau}, {Bourbeau}, {Bradascio}, {Braun}, {Bretz}, {Bron},
  {Brostean-Kaiser}, {Burgman}, {Busse}, {Carver}, {Chen}, {Cheung}, {Chirkin},
  {Clark}, {Classen}, {Collin}, {Conrad}, {Coppin}, {Correa}, {Cowen}, {Cross},
  {Dave}, {Day}, {de Andr{\'e}}, {De Clercq}, {DeLaunay}, {Dembinski},
  {Deoskar}, {De Ridder}, {Desiati}, {de Vries}, {de Wasseige}, {de With},
  {DeYoung}, {D{\'\i}az-V{\'e}lez}, {Dujmovic}, {Dunkman}, {Dvorak},
  {Eberhardt}, {Ehrhardt}, {Eichmann}, {Eller}, {Evenson}, {Fahey}, {Fazely},
  {Felde}, {Filimonov}, {Finley}, {Franckowiak}, {Friedman}, {Fritz},
  {Gaisser}, {Gallagher}, {Ganster}, {Garrappa}, {Gerhardt}, {Ghorbani},
  {Giang}, {Glauch}, {Gl{\"u}senkamp}, {Goldschmidt}, {Gonzalez}, {Grant},
  {Griffith}, {Haack}, {Hallgren}, {Halve}, {Halzen}, {Hanson}, {Hebecker},
  {Heereman}, {Helbing}, {Hellauer}, {Hickford}, {Hignight}, {Hill}, {Hoffman},
  {Hoffmann}, {Hoinka}, {Hokanson-Fasig}, {Hoshina}, {Huang}, {Huber},
  {Hultqvist}, {H{\"u}nnefeld}, {Hussain}, {In}, {Iovine}, {Ishihara},
  {Jacobi}, {Japaridze}, {Jeong}, {Jero}, {Jones}, {Kalaczynski}, {Kang},
  {Kappes}, {Kappesser}, {Karg}, {Karle}, {Katz}, {Kauer}, {Keivani}, {Kelley},
  {Kheirandish}, {Kim}, {Kintscher}, {Kiryluk}, {Kittler}, {Klein}, {Koirala},
  {Kolanoski}, {K{\"o}pke}, {Kopper}, {Kopper}, {Koskinen}, {Kowalski},
  {Krings}, {Kroll}, {Kr{\"u}ckl}, {Kunwar}, {Kurahashi}, {Kyriacou}, {Labare},
  {Lanfranchi}, {Larson}, {Lauber}, {Leonard}, {Leuermann}, {Liu}, {Lohfink},
  {Mariscal}, {Lu}, {L{\"u}nemann}, {Luszczak}, {Madsen}, {Maggi}, {Mahn},
  {Makino}, {Mancina}, {Mari{\textcommabelow s}}, {Maruyama}, {Mase}, {Maunu},
  {Meagher}, {Medici}, {Meier}, {Menne}, {Merino}, {Meures}, {Miarecki},
  {Micallef}, {Moment{\'e}}, {Montaruli}, {Moore}, {Moulai}, {Nagai},
  {Nahnhauer}, {Nakarmi}, {Naumann}, {Neer}, {Niederhausen}, {Nowicki},
  {Nygren}, {Obertacke Pollmann}, {Olivas}, {O'Murchadha}, {O'Sullivan},
  {Palczewski}, {Pandya}, {Pankova}, {Peiffer}, {P{\'e}rez de los Heros},
  {Pieloth}, {Pinat}, {Pizzuto}, {Plum}, {Price}, {Przybylski}, {Raab},
  {Rameez}, {Rauch}, {Rawlins}, {Rea}, {Reimann}, {Relethford}, {Renzi},
  {Resconi}, {Rhode}, {Richman}, {Robertson}, {Rongen}, {Rott}, {Ruhe},
  {Ryckbosch}, {Rysewyk}, {Safa}, {Sanchez Herrera}, {Sandrock}, {Sandroos},
  {Santander}, {Sarkar}, {Sarkar}, {Satalecka}, {Schaufel}, {Schlunder},
  {Schmidt}, {Schneider}, {Schneider}, {Sch{\"o}neberg}, {Schumacher},
  {Sclafani}, {Seckel}, {Seunarine}, {Soedingrekso}, {Soldin}, {Song},
  {Spiczak}, {Spiering}, {Stachurska}, {Stamatikos}, {Stanev}, {Stasik},
  {Stein}, {Stettner}, {Steuer}, {Stezelberger}, {Stokstad}, {St{\"o}{\ss}l},
  {Strotjohann}, {Stuttard}, {Sullivan}, {Sutherland}, {Taboada}, {Tenholt},
  {Ter-Antonyan}, {Terliuk}, {Tilav}, {Tobin}, {T{\"o}nnis}, {Toscano}, {Tosi},
  {Tselengidou}, {Tung}, {Turcati}, {Turcotte}, {Turley}, {Ty}, {Unger},
  {Unland Elorrieta}, {Usner}, {Vand enbroucke}, {Van Driessche}, {van Eijk},
  {van Eijndhoven}, {Vanheule}, {van Santen}, {Vraeghe}, {Walck}, {Wallace},
  {Wallraff}, {Wandler}, {Wandkowsky}, {Watson}, {Weaver}, {Weiss}, {Wendt},
  {Werthebach}, {Westerhoff}, {Whelan}, {Whitehorn}, {Wiebe}, {Wiebusch},
  {Wille}, {Williams}, {Wills}, {Wolf}, {Wood}, {Wood}, {Woolsey}, {Woschnagg},
  {Wrede}, {Xu}, {Xu}, {Xu}, {Yanez}, {Yodh}, {Yoshida}, \& {Yuan}}]{IceCube}
{Aartsen}, M.~G., {Ackermann}, M., {Adams}, J., {et~al.} 2019, European
  Physical Journal C, 79, 234, \dodoi{10.1140/epjc/s10052-019-6680-0}

\bibitem[{{Aartsen} {et~al.}(2021){Aartsen}, {Abbasi}, {Ackermann}, {Adams},
  {Aguilar}, {Ahlers}, {Ahrens}, {Alispach}, {Allison}, {Amin}, {Andeen},
  {Anderson}, {Ansseau}, {Anton}, {Arg{\"u}elles}, {Arlen}, {Auffenberg},
  {Axani}, {Bagherpour}, {Bai}, {Balagopal V.}, {Barbano}, {Bartos}, {Bastian},
  {Basu}, {Baum}, {Baur}, {Bay}, {Beatty}, {Becker}, {Becker Tjus}, {BenZvi},
  {Berley}, {Bernardini}, {Besson}, {Binder}, {Bindig}, {Blaufuss}, {Blot},
  {Bohm}, {Bohmer}, {B{\"o}ser}, {Botner}, {B{\"o}ttcher}, {Bourbeau},
  {Bourbeau}, {Bradascio}, {Braun}, {Bron}, {Brostean-Kaiser}, {Burgman},
  {Burley}, {Buscher}, {Busse}, {Bustamante}, {Campana}, {Carnie-Bronca},
  {Carver}, {Chen}, {Chen}, {Cheung}, {Chirkin}, {Choi}, {Clark}, {Clark},
  {Classen}, {Coleman}, {Collin}, {Connolly}, {Conrad}, {Coppin}, {Correa},
  {Cowen}, {Cross}, {Dave}, {Deaconu}, {De Clercq}, {DeLaunay}, {De Kockere},
  {Dembinski}, {Deoskar}, {De Ridder}, {Desai}, {Desiati}, {de Vries}, {de
  Wasseige}, {de With}, {DeYoung}, {Dharani}, {Diaz}, {D{\'\i}az-V{\'e}lez},
  {Dujmovic}, {Dunkman}, {DuVernois}, {Dvorak}, {Ehrhardt}, {Eller}, {Engel},
  {Evans}, {Evenson}, {Fahey}, {Farrag}, {Fazely}, {Felde}, {Fienberg},
  {Filimonov}, {Finley}, {Fischer}, {Fox}, {Franckowiak}, {Friedman}, {Fritz},
  {Gaisser}, {Gallagher}, {Ganster}, {Garcia-Fernand ez}, {Garrappa},
  {Gartner}, {Gerhardt}, {Gernhaeuser}, {Ghadimi}, {Glaser}, {Glauch},
  {Gl{\"u}senkamp}, {Goldschmidt}, {Gonzalez}, {Goswami}, {Grant},
  {Gr{\'e}goire}, {Griffith}, {Griswold}, {G{\"u}nd{\"u}z}, {Haack},
  {Hallgren}, {Halliday}, {Halve}, {Halzen}, {Hanson}, {Hanson}, {Hardin},
  {Haugen}, {Haungs}, {Hauser}, {Hebecker}, {Heinen}, {Heix}, {Helbing},
  {Hellauer}, {Henningsen}, {Hickford}, {Hignight}, {Hill}, {Hill}, {Hoffman},
  {Hoffmann}, {Hoffmann}, {Hoinka}, {Hokanson-Fasig}, {Holzapfel}, {Hoshina},
  {Huang}, {Huber}, {Huber}, {Huege}, {Hughes}, {Hultqvist}, {H{\"u}nnefeld},
  {Hussain}, {In}, {Iovine}, {Ishihara}, {Jansson}, {Japaridze}, {Jeong},
  {Jones}, {Jonske}, {Joppe}, {Kalekin}, {Kang}, {Kang}, {Kang}, {Kappes},
  {Kappesser}, {Karg}, {Karl}, {Karle}, {Katori}, {Katz}, {Kauer}, {Keivani},
  {Kellermann}, {Kelley}, {Kheirand ish}, {Kim}, {Kin}, {Kintscher}, {Kiryluk},
  {Kittler}, {Kleifges}, {Klein}, {Koirala}, {Kolanoski}, {K{\"o}pke},
  {Kopper}, {Kopper}, {Koskinen}, {Koundal}, {Kovacevich}, {Kowalski},
  {Krauss}, {Krings}, {Kr{\"u}ckl}, {Kulacz}, {Kurahashi}, {Lagunas Gualda},
  {Lahmann}, {Lanfranchi}, {Larson}, {Latif}, {Lauber}, {Lazar}, {Leonard},
  {Leszczy{\'n}ska}, {Li}, {Liu}, {Lohfink}, {LoSecco}, {Lozano Mariscal},
  {Lu}, {Lucarelli}, {Ludwig}, {L{\"u}nemann}, {Luszczak}, {Lyu}, {Ma},
  {Madsen}, {Maggi}, {Mahn}, {Makino}, {Mallik}, {Mancina}, {Mandalia},
  {Mari{\textcommabelow s}}, {Marka}, {Marka}, {Maruyama}, {Mase}, {Maunu},
  {McNally}, {Meagher}, {Medina}, {Meier}, {Meighen-Berger}, {Merz}, {Meyers},
  {Micallef}, {Mockler}, {Moment{\'e}}, {Montaruli}, {Moore}, {Morse},
  {Moulai}, {Muth}, {Naab}, {Nagai}, {Nam}, {Naumann}, {Necker}, {Neer},
  {Nelles}, {Nguy{\^e}n}, {Niederhausen}, {Nisa}, {Nowicki}, {Nygren},
  {Oberla}, {Obertacke Pollmann}, {Oehler}, {Olivas}, {O'Sullivan}, {Pan},
  {Pand ya}, {Pankova}, {Papp}, {Park}, {Parker}, {Paudel}, {Peiffer},
  {P{\'e}rez de los Heros}, {Petersen}, {Philippen}, {Pieloth}, {Pieper},
  {Pinfold}, {Pizzuto}, {Plaisier}, {Plum}, {Popovych}, {Porcelli}, {Prado
  Rodriguez}, {Price}, {Przybylski}, {Raab}, {Raissi}, {Rameez}, {Rauch},
  {Rawlins}, {Rea}, {Rehman}, {Reimann}, {Renschler}, {Renzi}, {Resconi},
  {Reusch}, {Rhode}, {Richman}, {Riedel}, {Riegel}, {Roberts}, {Robertson},
  {Roellinghoff}, {Rongen}, {Rott}, {Ruhe}, {Ryckbosch}, {Rysewyk Cantu},
  {Safa}, {Sanchez Herrera}, {Sand rock}, {Sandroos}, {Sandstrom}, {Santander},
  {Sarkar}, {Sarkar}, {Satalecka}, {Scharf}, {Schaufel}, {Schieler},
  {Schlunder}, {Schmidt}, {Schneider}, {Schneider}, {Schr{\"o}der},
  {Schumacher}, {Sclafani}, {Seckel}, {Seunarine}, {Shaevitz}, {Sharma},
  {Shefali}, {Silva}, {Smith}, {Smithers}, {Snihur}, {Soedingrekso}, {Soldin},
  {S{\"o}ldner-Rembold}, {Song}, {Southall}, {Spiczak}, {Spiering},
  {Stachurska}, {Stamatikos}, {Stanev}, {Stein}, {Stettner}, {Steuer},
  {Stezelberger}, {Stokstad}, {Strotjohann}, {St{\"u}rwald}, {Stuttard},
  {Sullivan}, {Taboada}, {Taketa}, {Tanaka}, {Tenholt}, {Ter-Antonyan},
  {Terliuk}, {Tilav}, {Tollefson}, {Tomankova}, {T{\"o}nnis}, {Torres},
  {Toscano}, {Tosi}, {Trettin}, {Tselengidou}, {Tung}, {Turcati}, {Turcotte},
  {Turley}, {Twagirayezu}, {Ty}, {Unger}, {Unland Elorrieta}, {Vand enbroucke},
  {van Eijk}, {van Eijndhoven}, {Vannerom}, {van Santen}, {Veberic},
  {Verpoest}, {Vieregg}, {Vraeghe}, {Walck}, {Watson}, {Weaver}, {Weindl},
  {Weinstock}, {Weiss}, {Weldert}, {Welling}, {Wendt}, {Werthebach},
  {Whitehorn}, {Wiebe}, {Wiebusch}, {Williams}, {Wissel}, {Wolf}, {Wood},
  {Woschnagg}, {Wrede}, {Wren}, {Wulff}, {Xu}, {Xu}, {Yanez}, {Yoshida},
  {Yuan}, {Zhang}, {Zierke}, \& {Z{\"o}cklein}}]{gen2}
{Aartsen}, M.~G., {Abbasi}, R., {Ackermann}, M., {et~al.} 2021, Journal of
  Physics G, 48, \dodoi{10.1088/1361-6471/abbd48}

\bibitem[{{Abdalla} {et~al.}(2018){Abdalla}, {Abramowski}, {Aharonian}, {Ait
  Benkhali}, {Ang{\"u}ner}, {Arakawa}, {Arrieta}, {Aubert}, {Backes}, {Balzer},
  {Barnard}, {Becherini}, {Becker Tjus}, {Berge}, {Bernhard}, {Bernl{\"o}hr},
  {Blackwell}, {B{\"o}ttcher}, {Boisson}, {Bolmont}, {Bonnefoy}, {Bordas},
  {Bregeon}, {Brun}, {Brun}, {Bryan}, {B{\"u}chele}, {Bulik}, {Capasso},
  {Carrigan}, {Caroff}, {Carosi}, {Casanova}, {Cerruti}, {Chakraborty},
  {Chaves}, {Chen}, {Chevalier}, {Colafrancesco}, {Condon}, {Conrad}, {Davids},
  {Decock}, {Deil}, {Devin}, {deWilt}, {Dirson}, {Djannati-Ata{\"\i}},
  {Domainko}, {Donath}, {Drury}, {Dutson}, {Dyks}, {Edwards}, {Egberts},
  {Eger}, {Emery}, {Ernenwein}, {Eschbach}, {Farnier}, {Fegan}, {Fernandes},
  {Fiasson}, {Fontaine}, {F{\"o}rster}, {Funk}, {F{\"u}{\ss}ling}, {Gabici},
  {Gallant}, {Garrigoux}, {Gast}, {Gat{\'e}}, {Giavitto}, {Giebels}, {Glawion},
  {Glicenstein}, {Gottschall}, {Grondin}, {Hahn}, {Haupt}, {Hawkes},
  {Heinzelmann}, {Henri}, {Hermann}, {Hinton}, {Hofmann}, {Hoischen}, {Holch},
  {Holler}, {Horns}, {Ivascenko}, {Iwasaki}, {Jacholkowska}, {Jamrozy},
  {Jankowsky}, {Jankowsky}, {Jingo}, {Jouvin}, {Jung-Richardt}, {Kastendieck},
  {Katarzy{\'n}ski}, {Katsuragawa}, {Katz}, {Kerszberg}, {Khangulyan},
  {Kh{\'e}lifi}, {King}, {Klepser}, {Klochkov}, {Klu{\'z}niak}, {Komin},
  {Kosack}, {Krakau}, {Kraus}, {Kr{\"u}ger}, {Laffon}, {Lamanna}, {Lau},
  {Lees}, {Lefaucheur}, {Lemi{\`e}re}, {Lemoine-Goumard}, {Lenain}, {Leser},
  {Lohse}, {Lorentz}, {Liu}, {L{\'o}pez-Coto}, {Lypova}, {Marandon},
  {Malyshev}, {Marcowith}, {Mariaud}, {Marx}, {Maurin}, {Maxted}, {Mayer},
  {Meintjes}, {Meyer}, {Mitchell}, {Moderski}, {Mohamed}, {Mohrmann},
  {Mor{\r{a}}}, {Moulin}, {Murach}, {Nakashima}, {de Naurois}, {Ndiyavala},
  {Niederwanger}, {Niemiec}, {Oakes}, {O'Brien}, {Odaka}, {Ohm}, {Ostrowski},
  {Oya}, {Padovani}, {Panter}, {Parsons}, {Paz Arribas}, {Pekeur}, {Pelletier},
  {Perennes}, {Petrucci}, {Peyaud}, {Piel}, {Pita}, {Poireau}, {Poon},
  {Prokhorov}, {Prokoph}, {P{\"u}hlhofer}, {Punch}, {Quirrenbach}, {Raab},
  {Rauth}, {Reimer}, {Reimer}, {Renaud}, {de los Reyes}, {Rieger}, {Rinchiuso},
  {Romoli}, {Rowell}, {Rudak}, {Rulten}, {Safi-Harb}, {Sahakian}, {Saito},
  {Sanchez}, {Santangelo}, {Sasaki}, {Schandri}, {Schlickeiser},
  {Sch{\"u}ssler}, {Schulz}, {Schwanke}, {Schwemmer}, {Seglar-Arroyo},
  {Settimo}, {Seyffert}, {Shafi}, {Shilon}, {Shiningayamwe}, {Simoni}, {Sol},
  {Spanier}, {Spir-Jacob}, {Stawarz}, {Steenkamp}, {Stegmann}, {Steppa},
  {Sushch}, {Takahashi}, {Tavernet}, {Tavernier}, {Taylor}, {Terrier},
  {Tibaldo}, {Tiziani}, {Tluczykont}, {Trichard}, {Tsirou}, {Tsuji}, {Tuffs},
  {Uchiyama}, {van der Walt}, {van Eldik}, {van Rensburg}, {van Soelen},
  {Vasileiadis}, {Veh}, {Venter}, {Viana}, {Vincent}, {Vink}, {Voisin},
  {V{\"o}lk}, {Vuillaume}, {Wadiasingh}, {Wagner}, {Wagner}, {Wagner}, {White},
  {Wierzcholska}, {Willmann}, {W{\"o}rnlein}, {Wouters}, {Yang}, {Zaborov},
  {Zacharias}, {Zanin}, {Zdziarski}, {Zech}, {Zefi}, {Ziegler}, {Zorn}, \&
  {{\.Z}ywucka}}]{HGPS}
{Abdalla}, H., {Abramowski}, A., {Aharonian}, F., {et~al.} 2018, \aap, 612, A1,
  \dodoi{10.1051/0004-6361/201732098}

\bibitem[{{Abdalla} {et~al.}(2021){Abdalla}, {Aharonian}, {Ait Benkhali},
  {Ang{\"u}ner}, {Arcaro}, {Armand}, {Armstrong}, {Ashkar}, {Backes},
  {Baghmanyan}, {Barbosa Martins}, {Barnacka}, {Barnard}, {Becherini}, {Berge},
  {Bernl{\"o}hr}, {Bi}, {B{\"o}ttcher}, {Boisson}, {Bolmont}, {de Bonyde
  Lavergne}, {Breuhaus}, {Brose}, {Brun}, {Brun}, {Bryan}, {B{\"u}chele},
  {Bulik}, {Bylund}, {Caroff}, {Carosi}, {Chand}, {Chandra}, {Chen}, {Cotter},
  {Cury{\l}o}, {Damascene Mbarubucyeye}, {Davids}, {Davies}, {Deil}, {Devin},
  {Dirson}, {Djannati-Ata{\"\i}}, {Dmytriiev}, {Donath}, {Doroshenko},
  {Dreyer}, {Duffy}, {Dyks}, {Egberts}, {Eichhorn}, {Einecke}, {Emery},
  {Ernenwein}, {Feijen}, {Fegan}, {Fiasson}, {Fichet de Clairfontaine},
  {Fontaine}, {Funk}, {F{\"u}{\ss}ling}, {Gabici}, {Gallant}, {Giavitto},
  {Giunti}, {Glawion}, {Glicenstein}, {Gottschall}, {Grondin}, {Hahn}, {Haupt},
  {Hermann}, {Hinton}, {Hofmann}, {Hoischen}, {Holch}, {Holler}, {H{\"o}rbe},
  {Horns}, {Huber}, {Jamrozy}, {Jankowsky}, {Jankowsky}, {Jung-Richardt},
  {Kasai}, {Kastendieck}, {Katarzy{\'n}ski}, {Katz}, {Khangulyan},
  {Kh{\'e}lifi}, {Klepser}, {Klu{\'z}niak}, {Komin}, {Konno}, {Kosack},
  {Kostunin}, {Kreter}, {Lamanna}, {Lemi{\`e}re}, {Lemoine-Goumard}, {Lenain},
  {Leuschner}, {Levy}, {Lohse}, {Lypova}, {Mackey}, {Majumdar}, {Malyshev},
  {Malyshev}, {Marandon}, {Marchegiani}, {Marcowith}, {Mares},
  {Mart{\'\i}-Devesa}, {Marx}, {Maurin}, {Meintjes}, {Meyer}, {Mitchell},
  {Moderski}, {Mohrmann}, {Montanari}, {Moore}, {Morris}, {Moulin}, {Muller},
  {Murach}, {Nakashima}, {Nayerhoda}, {de Naurois}, {Ndiyavala}, {Niemiec},
  {Oakes}, {O'Brien}, {Odaka}, {Ohm}, {Olivera-Nieto}, {de Ona Wilhelmi},
  {Ostrowski}, {Panter}, {Panny}, {Parsons}, {Peron}, {Peyaud}, {Piel}, {Pita},
  {Poireau}, {Priyana Noel}, {Prokhorov}, {Prokoph}, {P{\"u}hlhofer}, {Punch},
  {Quirrenbach}, {Raab}, {Rauth}, {Reichherzer}, {Reimer}, {Reimer}, {Remy},
  {Renaud}, {Rieger}, {Rinchiuso}, {Romoli}, {Rowell}, {Rudak}, {Sahakian},
  {Sailer}, {Salzmann}, {Sanchez}, {Santangelo}, {Sasaki}, {Sch{\"a}fer},
  {Sch{\"u}ssler}, {Schutte}, {Schwanke}, {Seglar-Arroyo}, {Senniappan},
  {Seyffert}, {Shafi}, {Shapopi}, {Shiningayamwe}, {Simoni}, {Sinha}, {Sol},
  {Specovius}, {Spencer}, {Spir-Jacob}, {Stawarz}, {Sun}, {Steenkamp},
  {Stegmann}, {Steinmassl}, {Steppa}, {Takahashi}, {Tavernier}, {Taylor},
  {Terrier}, {Thiersen}, {Tiziani}, {Tluczykont}, {Tomankova}, {Trichard},
  {Tsirou}, {Tuffs}, {Uchiyama}, {van der Walt}, {van Eldik}, {van Rensburg},
  {van Soelen}, {Vasileiadis}, {Veh}, {Venter}, {Vincent}, {Vink}, {V{\"o}lk},
  {Wadiasingh}, {Wagner}, {Watson}, {Werner}, {White}, {Wierzcholska}, {Wong},
  {Yusafzai}, {Zacharias}, {Zanin}, {Zargaryan}, {Zdziarski}, {Zech}, {Zhu},
  {Zmija}, {Zorn}, {Zouari}, {{\.Z}ywucka}, {Albert}, {Alfaro}, {Alvarez},
  {Arteaga-Vel{\'e}zquez}, {Arunbabu}, {Avila Rojas}, {Belmont-Moreno},
  {BenZvi}, {Brisbois}, {Caballero-Mora}, {Capistr{\'a}n}, {Carrami{\~n}ana},
  {Casanova}, {Cotti}, {Cotzomi}, {Couti{\~n}o de Le{\'o}n}, {De la Fuente},
  {de Le{\'o}n}, {Diaz Hernandez}, {D{\'\i}az-V{\'e}lez}, {Dingus},
  {DuVernois}, {Durocher}, {Ellsworth}, {Engel}, {Espinoza}, {Fan},
  {Fern{\'a}ndez Alonso}, {Fraija}, {Galv{\'a}n-G{\'a}mez}, {Garcia},
  {Garc{\'\i}a-Gonz{\'a}lez}, {Garfias}, {Giacinti}, {Gonz{\'a}lez}, {Goodman},
  {Harding}, {Hernandez}, {Hona}, {Huang}, {Hueyotl-Zahuantitla},
  {H{\"u}ntemeyer}, {Iriarte}, {Jardin-Blicq}, {Joshi}, {Kieda}, {Lee},
  {Le{\'o}n Vargas}, {Linnemann}, {Longinotti}, {Luis-Raya}, {L{\'o}pez-Coto},
  {Malone}, {Martinez}, {Martinez-Castellanos}, {Mart{\'\i}nez-Castro},
  {Matthews}, {Miranda-Romagnoli}, {Morales-Soto}, {Moreno}, {Mostaf{\'a}},
  {Nayerhoda}, {Nellen}, {Newbold}, {Nisa}, {Noriega-Papaqui}, {Omodei},
  {Peisker}, {P{\'e}rez Araujo}, {P{\'e}rez-P{\'e}rez}, {Rho},
  {Rosa-Gonz{\'a}lez}, {Ruiz-Velasco}, {Salesa Greus}, {Sandoval}, {Schneider},
  {Schoorlemmer}, {Serna-Franco}, {Smith}, {Springer}, {Surajbali},
  {Tollefson}, {Torres}, {Torres-Escobedo}, {Turner}, {Ure{\~n}a-Mena},
  {Villase{\~n}or}, {Weisgarber}, {Willox}, {Zhou}, \& {HAWC
  Collaboration}}]{2021ApJ...917....6A}
{Abdalla}, H., {Aharonian}, F., {Ait Benkhali}, F., {et~al.} 2021, \apj, 917,
  6, \dodoi{10.3847/1538-4357/abf64b}

\bibitem[{{Abdo} {et~al.}(2007){Abdo}, {Allen}, {Berley}, {Casanova}, {Chen},
  {Coyne}, {Dingus}, {Ellsworth}, {Fleysher}, {Fleysher}, {Gonzalez},
  {Goodman}, {Hays}, {Hoffman}, {Hopper}, {H{\"u}ntemeyer}, {Kolterman},
  {Lansdell}, {Linnemann}, {McEnery}, {Mincer}, {Nemethy}, {Noyes}, {Ryan},
  {Saz Parkinson}, {Shoup}, {Sinnis}, {Smith}, {Sullivan}, {Vasileiou},
  {Walker}, {Williams}, {Xu}, \& {Yodh}}]{Milagro}
{Abdo}, A.~A., {Allen}, B., {Berley}, D., {et~al.} 2007, Astrophysical Journal
  Letters, 664, L91, \dodoi{10.1086/520717}

\bibitem[{{Abdo} {et~al.}(2008){Abdo}, {Allen}, {Aune}, {Berley}, {Blaufuss},
  {Casanova}, {Chen}, {Dingus}, {Ellsworth}, {Fleysher}, {Fleysher},
  {Gonzalez}, {Goodman}, {Hoffman}, {H{\"u}ntemeyer}, {Kolterman}, {Lansdell},
  {Linnemann}, {McEnery}, {Mincer}, {Moskalenko}, {Nemethy}, {Noyes}, {Porter},
  {Pretz}, {Ryan}, {Saz Parkinson}, {Shoup}, {Sinnis}, {Smith}, {Strong},
  {Sullivan}, {Vasileiou}, {Walker}, {Williams}, \& {Yodh}}]{diffuse}
{Abdo}, A.~A., {Allen}, B., {Aune}, T., {et~al.} 2008, The Astrophysical
  Journal, 688, 1078, \dodoi{10.1086/592213}

\bibitem[{{Abdo} {et~al.}(2010){Abdo}, {Ackermann}, {Ajello}, {Baldini},
  {Ballet}, {Barbiellini}, {Bastieri}, {Baughman}, {Bechtol}, {Bellazzini},
  {Berenji}, {Blandford}, {Bloom}, {Bonamente}, {Borgland}, {Bregeon}, {Brez},
  {Brigida}, {Bruel}, {Burnett}, {Buson}, {Caliandro}, {Cameron}, {Camilo},
  {Caraveo}, {Casandjian}, {Cecchi}, {{\c{C}}elik}, {Chekhtman}, {Cheung},
  {Chiang}, {Ciprini}, {Claus}, {Cognard}, {Cohen-Tanugi}, {Cominsky},
  {Conrad}, {Cutini}, {de Angelis}, {de Palma}, {Digel}, {Dingus}, {Dormody},
  {Silva}, {Drell}, {Dubois}, {Dumora}, {Farnier}, {Favuzzi}, {Fegan}, {Focke},
  {Fortin}, {Frailis}, {Freire}, {Fukazawa}, {Funk}, {Fusco}, {Gargano},
  {Gasparrini}, {Gehrels}, {Germani}, {Giavitto}, {Giebels}, {Giglietto},
  {Giordano}, {Glanzman}, {Godfrey}, {Grenier}, {Grondin}, {Grove},
  {Guillemot}, {Guiriec}, {Hanabata}, {Harding}, {Hays}, {Hughes}, {Jackson},
  {J{\'o}hannesson}, {Johnson}, {Johnson}, {Johnson}, {Johnston}, {Kamae},
  {Katagiri}, {Kataoka}, {Kawai}, {Kerr}, {Kn{\"o}dlseder}, {Kocian}, {Kuss},
  {Lande}, {Latronico}, {Lemoine-Goumard}, {Longo}, {Loparco}, {Lott},
  {Lovellette}, {Lubrano}, {Makeev}, {Marelli}, {Mazziotta}, {McEnery},
  {Meurer}, {Michelson}, {Mitthumsiri}, {Mizuno}, {Moiseev}, {Monte},
  {Monzani}, {Morselli}, {Moskalenko}, {Murgia}, {Nolan}, {Norris}, {Nuss},
  {Ohsugi}, {Omodei}, {Orlando}, {Ormes}, {Paneque}, {Parent}, {Pelassa},
  {Pepe}, {Pesce-Rollins}, {Piron}, {Porter}, {Rain{\`o}}, {Rando}, {Ray},
  {Razzano}, {Reimer}, {Reimer}, {Reposeur}, {Ritz}, {Roberts}, {Rochester},
  {Rodriguez}, {Ro'mani}, {Roth}, {Ryde}, {Sadrozinski}, {Sanchez}, {Sander},
  {Saz Parkinson}, {Scargle}, {Sgr{\`o}}, {Siskind}, {Smith}, {Smith},
  {Spandre}, {Spinelli}, {Strickman}, {Suson}, {Tajima}, {Takahashi}, {Tanaka},
  {Thayer}, {Thayer}, {Theureau}, {Thompson}, {Tibaldo}, {Tibolla}, {Torres},
  {Tosti}, {Tramacere}, {Uchiyama}, {Usher}, {Van Etten}, {Vasileiou},
  {Venter}, {Vilchez}, {Vitale}, {Waite}, {Wang}, {Watters}, {Winer}, {Wolff},
  {Wood}, {Ylinen}, \& {Ziegler}}]{Fermi}
{Abdo}, A.~A., {Ackermann}, M., {Ajello}, M., {et~al.} 2010, The Astrophysical
  Journal, 711, 64, \dodoi{10.1088/0004-637X/711/1/64}

\bibitem[{Abeysekara {et~al.}(2020)Abeysekara, Albert, Alfaro,
  {et~al.}}]{HECatalog}
Abeysekara, A., Albert, A., Alfaro, R., {et~al.} 2020, Phys. Rev. Lett., 124,
  021102, \dodoi{10.1103/PhysRevLett.124.021102}

\bibitem[{{Abeysekara} {et~al.}(2017{\natexlab{a}}){Abeysekara}, {Albert},
  {Alfaro}, {Alvarez}, {{\'A}lvarez}, {Arceo}, {Arteaga-Vel{\'a}zquez}, {Ayala
  Solares}, {Barber}, {Bautista-Elivar}, {Becerril}, {Belmont-Moreno},
  {BenZvi}, {Berley}, {Braun}, {Brisbois}, {Caballero-Mora}, {Capistr{\'a}n},
  {Carrami{\~n}ana}, {Casanova}, {Castillo}, {Cotti}, {Cotzomi}, {Couti{\~n}o
  de Le{\'o}n}, {de la Fuente}, {De Le{\'o}n}, {DeYoung}, {Dingus},
  {DuVernois}, {D{\'\i}az-V{\'e}lez}, {Ellsworth}, {Fiorino}, {Fraija},
  {Garc{\'\i}a-Gonz{\'a}lez}, {Gerhardt}, {Gonz{\'a}lez Mun{\"o}z},
  {Gonz{\'a}lez}, {Goodman}, {Hampel-Arias}, {Harding}, {Hernandez},
  {Hernandez-Almada}, {Hinton}, {Hui}, {H{\"u}ntemeyer}, {Iriarte},
  {Jardin-Blicq}, {Joshi}, {Kaufmann}, {Kieda}, {Lara}, {Lauer}, {Lee},
  {Lennarz}, {Le{\'o}n Vargas}, {Linnemann}, {Longinotti}, {Raya},
  {Luna-Garc{\'\i}a}, {L{\'o}pez-Coto}, {Malone}, {Marinelli}, {Martinez},
  {Martinez-Castellanos}, {Mart{\'\i}nez-Castro}, {Mart{\'\i}nez-Huerta},
  {Matthews}, {Miranda-Romagnoli}, {Moreno}, {Mostaf{\'a}}, {Nellen},
  {Newbold}, {Nisa}, {Noriega-Papaqui}, {Pelayo}, {Pretz},
  {P{\'e}rez-P{\'e}rez}, {Ren}, {Rho}, {Rivi{\`e}re}, {Rosa-Gonz{\'a}lez},
  {Rosenberg}, {Ruiz-Velasco}, {Salazar}, {Salesa Greus}, {Sandoval},
  {Schneider}, {Schoorlemmer}, {Sinnis}, {Smith}, {Springer}, {Surajbali},
  {Taboada}, {Tibolla}, {Tollefson}, {Torres}, {Ukwatta}, {Villase{\~n}or},
  {Weisgarber}, {Westerhoff}, {Wisher}, {Wood}, {Yapici}, {Yodh}, {Younk},
  {Zepeda}, \& {Zhou}}]{Crab2017}
{Abeysekara}, A.~U., {Albert}, A., {Alfaro}, R., {et~al.} 2017{\natexlab{a}},
  \apj, 843, 39, \dodoi{10.3847/1538-4357/aa7555}

\bibitem[{{Abeysekara} {et~al.}(2017{\natexlab{b}}){Abeysekara}, {Albert},
  {Alfaro}, {Alvarez}, {{\'A}lvarez}, {Arceo}, {Arteaga-Vel{\'a}zquez}, {Avila
  Rojas}, {Ayala Solares}, {Barber}, {Bautista-Elivar}, {Becerril},
  {Belmont-Moreno}, {BenZvi}, {Berley}, {Bernal}, {Braun}, {Brisbois},
  {Caballero-Mora}, {Capistr{\'a}n}, {Carrami{\~n}ana}, {Casanova}, {Castillo},
  {Cotti}, {Cotzomi}, {Couti{\~n}o de Le{\'o}n}, {De Le{\'o}n}, {De la Fuente},
  {Dingus}, {DuVernois}, {D{\'\i}az-V{\'e}lez}, {Ellsworth}, {Engel},
  {Enr{\'\i}quez-Rivera}, {Fiorino}, {Fraija}, {Garc{\'\i}a-Gonz{\'a}lez},
  {Garfias}, {Gerhardt}, {Gonz{\'a}lez Mu{\~n}oz}, {Gonz{\'a}lez}, {Goodman},
  {Hampel-Arias}, {Harding}, {Hern{\'a}ndez}, {Hern{\'a}ndez-Almada}, {Hinton},
  {Hona}, {Hui}, {H{\"u}ntemeyer}, {Iriarte}, {Jardin-Blicq}, {Joshi},
  {Kaufmann}, {Kieda}, {Lara}, {Lauer}, {Lee}, {Lennarz}, {Vargas},
  {Linnemann}, {Longinotti}, {Luis Raya}, {Luna-Garc{\'\i}a}, {L{\'o}pez-Coto},
  {Malone}, {Marinelli}, {Martinez}, {Martinez-Castellanos},
  {Mart{\'\i}nez-Castro}, {Mart{\'\i}nez-Huerta}, {Matthews}, {Mirand
  a-Romagnoli}, {Moreno}, {Mostaf{\'a}}, {Nellen}, {Newbold}, {Nisa},
  {Noriega-Papaqui}, {Pelayo}, {Pretz}, {P{\'e}rez-P{\'e}rez}, {Ren}, {Rho},
  {Rivi{\`e}re}, {Rosa-Gonz{\'a}lez}, {Rosenberg}, {Ruiz-Velasco}, {Salazar},
  {Salesa Greus}, {Sand oval}, {Schneider}, {Schoorlemmer}, {Sinnis}, {Smith},
  {Springer}, {Surajbali}, {Taboada}, {Tibolla}, {Tollefson}, {Torres},
  {Ukwatta}, {Vianello}, {Weisgarber}, {Westerhoff}, {Wisher}, {Wood},
  {Yapici}, {Yodh}, {Younk}, {Zepeda}, {Zhou}, {Guo}, {Hahn}, {Li}, \&
  {Zhang}}]{Geminga}
---. 2017{\natexlab{b}}, Science, 358, 911, \dodoi{10.1126/science.aan4880}

\bibitem[{{Abeysekara} {et~al.}(2017{\natexlab{c}}){Abeysekara}, {Albert},
  {Alfaro}, {Alvarez}, {{\'A}lvarez}, {Arceo}, {Arteaga-Vel{\'a}zquez}, {Ayala
  Solares}, {Barber}, {Baughman}, {Bautista-Elivar}, {Becerra Gonzalez},
  {Becerril}, {Belmont-Moreno}, {BenZvi}, {Berley}, {Bernal}, {Braun},
  {Brisbois}, {Caballero-Mora}, {Capistr{\'a}n}, {Carrami{\~n}ana}, {Casanova},
  {Castillo}, {Cotti}, {Cotzomi}, {Couti{\~n}o de Le{\'o}n}, {de la Fuente},
  {De Le{\'o}n}, {Diaz Hernandez}, {Dingus}, {DuVernois},
  {D{\'\i}az-V{\'e}lez}, {Ellsworth}, {Engel}, {Fiorino}, {Fraija},
  {Garc{\'\i}a-Gonz{\'a}lez}, {Garfias}, {Gerhardt}, {Gonz{\'a}lez Mu{\~n}oz},
  {Gonz{\'a}lez}, {Goodman}, {Hampel-Arias}, {Harding}, {Hernandez},
  {Hernandez-Almada}, {Hinton}, {Hui}, {H{\"u}ntemeyer}, {Iriarte},
  {Jardin-Blicq}, {Joshi}, {Kaufmann}, {Kieda}, {Lara}, {Lauer}, {Lee},
  {Lennarz}, {Le{\'o}n Vargas}, {Linnemann}, {Longinotti}, {Raya},
  {Luna-Garc{\'\i}a}, {L{\'o}pez-Coto}, {Malone}, {Marinelli}, {Martinez},
  {Martinez-Castellanos}, {Mart{\'\i}nez-Castro}, {Mart{\'\i}nez-Huerta},
  {Matthews}, {Miranda-Romagnoli}, {Moreno}, {Mostaf{\'a}}, {Nellen},
  {Newbold}, {Nisa}, {Noriega-Papaqui}, {Pelayo}, {Pretz},
  {P{\'e}rez-P{\'e}rez}, {Ren}, {Rho}, {Rivi{\`e}re}, {Rosa-Gonz{\'a}lez},
  {Rosenberg}, {Ruiz-Velasco}, {Salazar}, {Salesa Greus}, {Sandoval},
  {Schneider}, {Schoorlemmer}, {Sinnis}, {Smith}, {Springer}, {Surajbali},
  {Taboada}, {Tibolla}, {Tollefson}, {Torres}, {Ukwatta}, {Vianello},
  {Villase{\~n}or}, {Weisgarber}, {Westerhoff}, {Wisher}, {Wood}, {Yapici},
  {Younk}, {Zepeda}, \& {Zhou}}]{2hwc}
---. 2017{\natexlab{c}}, The Astrophysical Journal, 843, 40,
  \dodoi{10.3847/1538-4357/aa7556}

\bibitem[{{Abeysekara} {et~al.}(2018){Abeysekara}, {Albert}, {Alfaro},
  {Alvarez}, {{\'A}lvarez}, {Arceo}, {Arteaga-Vel{\'a}zquez}, {Avila Rojas},
  {Ayala Solares}, {Belmont-Moreno}, {BenZvi}, {Brisbois}, {Caballero-Mora},
  {Capistr{\'a}n}, {Carrami{\~n}ana}, {Casanova}, {Castillo}, {Cotti},
  {Cotzomi}, {Couti{\~n}o de Le{\'o}n}, {De Le{\'o}n}, {De la Fuente},
  {D{\'\i}az-V{\'e}lez}, {Dichiara}, {Dingus}, {DuVernois}, {Ellsworth},
  {Engel}, {Espinoza}, {Fang}, {Fleischhack}, {Fraija}, {Galv{\'a}n-G{\'a}mez},
  {Garc{\'\i}a-Gonz{\'a}lez}, {Garfias}, {Gonz{\'a}lez-Mu{\~n}oz},
  {Gonz{\'a}lez}, {Goodman}, {Hampel-Arias}, {Harding}, {Hernandez}, {Hinton},
  {Hona}, {Hueyotl-Zahuantitla}, {Hui}, {H{\"u}ntemeyer}, {Iriarte},
  {Jardin-Blicq}, {Joshi}, {Kaufmann}, {Kar}, {Kunde}, {Lauer}, {Lee},
  {Le{\'o}n Vargas}, {Li}, {Linnemann}, {Longinotti}, {Luis-Raya},
  {L{\'o}pez-Coto}, {Malone}, {Marinelli}, {Martinez}, {Martinez-Castellanos},
  {Mart{\'\i}nez-Castro}, {Matthews}, {Mirand a-Romagnoli}, {Moreno},
  {Mostaf{\'a}}, {Nayerhoda}, {Nellen}, {Newbold}, {Nisa}, {Noriega-Papaqui},
  {Pretz}, {P{\'e}rez-P{\'e}rez}, {Ren}, {Rho}, {Rivi{\`e}re},
  {Rosa-Gonz{\'a}lez}, {Rosenberg}, {Ruiz-Velasco}, {Salesa Greus}, {Sandoval},
  {Schneider}, {Schoorlemmer}, {Seglar Arroyo}, {Sinnis}, {Smith}, {Springer},
  {Surajbali}, {Taboada}, {Tibolla}, {Tollefson}, {Torres}, {Vianello},
  {Villase{\~n}or}, {Weisgarber}, {Werner}, {Westerhoff}, {Wood}, {Yapici},
  {Yodh}, {Zepeda}, {Zhang}, \& {Zhou}}]{ss433}
---. 2018, Nature, 562, 82, \dodoi{10.1038/s41586-018-0565-5}

\bibitem[{{Abeysekara} {et~al.}(2019){Abeysekara}, {Albert}, {Alfaro},
  {Alvarez}, {{\'A}lvarez}, {Camacho}, {Arceo}, {Arteaga-Vel{\'a}zquez},
  {Arunbabu}, {Avila Rojas}, {Ayala Solares}, {Baghmanyan}, {Belmont-Moreno},
  {BenZvi}, {Brisbois}, {Caballero-Mora}, {Capistr{\'a}n}, {Carrami{\~n}ana},
  {Casanova}, {Cotti}, {Cotzomi}, {Couti{\~n}o de Le{\'o}n}, {De la Fuente},
  {de Le{\'o}n}, {Dichiara}, {Dingus}, {DuVernois}, {D{\'\i}az-V{\'e}lez},
  {Ellsworth}, {Engel}, {Espinoza}, {Fick}, {Fleischhack}, {Fraija},
  {Galv{\'a}n-G{\'a}mez}, {Garc{\'\i}a-Gonz{\'a}lez}, {Garfias},
  {Gonz{\'a}lez}, {Goodman}, {Harding}, {Hernandez}, {Hinton}, {Hona},
  {Hueyotl-Zahuantitla}, {Hui}, {H{\"u}ntemeyer}, {Iriarte}, {Jardin-Blicq},
  {Joshi}, {Kaufmann}, {Kieda}, {Lara}, {Lee}, {Le{\'o}n Vargas}, {Linnemann},
  {Longinotti}, {Luis-Raya}, {Lundeen}, {Malone}, {Marinelli}, {Martinez},
  {Martinez-Castellanos}, {Mart{\'\i}nez-Castro}, {Mart{\'\i}nez-Huerta},
  {Matthews}, {Miranda-Romagnoli}, {Morales-Soto}, {Moreno}, {Mostaf{\'a}},
  {Nayerhoda}, {Nellen}, {Newbold}, {Nisa}, {Noriega-Papaqui}, {Peisker},
  {P{\'e}rez-P{\'e}rez}, {Pretz}, {Ren}, {Rho}, {Rivi{\`e}re},
  {Rosa-Gonz{\'a}lez}, {Rosenberg}, {Ruiz-Velasco}, {Salazar}, {Salesa Greus},
  {Sandoval}, {Schneider}, {Schoorlemmer}, {Seglar Arroyo}, {Sinnis}, {Smith},
  {Springer}, {Surajbali}, {Tabachnick}, {Tanner}, {Tibolla}, {Tollefson},
  {Torres}, {Weisgarber}, {Westerhoff}, {Wood}, {Yapici}, {Zepeda}, {Zhou}, \&
  {HAWC Collaboration}}]{Crab}
---. 2019, The Astrophysical Journal, 881, 134,
  \dodoi{10.3847/1538-4357/ab2f7d}

\bibitem[{{Abeysekara} {et~al.}(2020){Abeysekara}, {Albert}, {Alfaro}, {Angeles
  Camacho}, {Arteaga-Vel{\'a}zquez}, {Arunbabu}, {Avila Rojas}, {Ayala
  Solares}, {Baghmanyan}, {Belmont-Moreno}, {BenZvi}, {Brisbois},
  {Caballero-Mora}, {Capistr{\'a}n}, {Carrami{\~n}ana}, {Casanova}, {Cotti},
  {Cotzomi}, {Couti{\~n}o de Le{\'o}n}, {De la Fuente}, {de Le{\'o}n},
  {Dichiara}, {Dingus}, {DuVernois}, {D{\'\i}az-V{\'e}lez}, {Ellsworth},
  {Engel}, {Espinoza}, {Fleischhack}, {Fraija}, {Galv{\'a}n-G{\'a}mez},
  {Garcia}, {Garc{\'\i}a-Gonz{\'a}lez}, {Garfias}, {Gonz{\'a}lez}, {Goodman},
  {Harding}, {Hernandez}, {Hinton}, {Hona}, {Huang}, {Hueyotl-Zahuantitla},
  {H{\"u}ntemeyer}, {Iriarte}, {Jardin-Blicq}, {Joshi}, {Kaufmann}, {Kieda},
  {Lara}, {Lee}, {Le{\'o}n Vargas}, {Linnemann}, {Longinotti}, {Luis-Raya},
  {Lundeen}, {L{\'o}pez-Coto}, {Malone}, {Marinelli}, {Martinez},
  {Martinez-Castellanos}, {Mart{\'\i}nez-Castro}, {Mart{\'\i}nez-Huerta},
  {Matthews}, {Miranda-Romagnoli}, {Morales-Soto}, {Moreno}, {Mostaf{\'a}},
  {Nayerhoda}, {Nellen}, {Newbold}, {Nisa}, {Noriega-Papaqui}, {Peisker},
  {P{\'e}rez-P{\'e}rez}, {Pretz}, {Ren}, {Rho}, {Rivi{\`e}re},
  {Rosa-Gonz{\'a}lez}, {Rosenberg}, {Ruiz-Velasco}, {Salesa Greus}, {Sand
  oval}, {Schneider}, {Schoorlemmer}, {Sinnis}, {Smith}, {Springer},
  {Surajbali}, {Tabachnick}, {Tanner}, {Tibolla}, {Tollefson}, {Torres},
  {Torres-Escobedo}, {Villase{\~n}or}, {Weisgarber}, {Wood}, {Yapici}, {Zhang},
  {Zhou}, \& {HAWC Collaboration}}]{catalog}
---. 2020, Physical Review Letters, 124, 021102,
  \dodoi{10.1103/PhysRevLett.124.021102}

\bibitem[{{Aharonian} {et~al.}(2009){Aharonian}, {Akhperjanian}, {Anton},
  {Barres de Almeida}, {Bazer-Bachi}, {Becherini}, {Behera}, {Benbow},
  {Bernl{\"o}hr}, {Boisson}, {Bochow}, {Borrel}, {Braun}, {Brion}, {Brucker},
  {Brun}, {B{\"u}hler}, {Bulik}, {B{\"u}sching}, {Boutelier}, {Carrigan},
  {Chadwick}, {Charbonnier}, {Chaves}, {Cheesebrough}, {Chounet}, {Clapson},
  {Coignet}, {Dalton}, {Daniel}, {Degrange}, {Deil}, {Dickinson},
  {Djannati-Ata{\"\i}}, {Domainko}, {O'C. Drury}, {Dubois}, {Dubus}, {Dyks},
  {Dyrda}, {Egberts}, {Emmanoulopoulos}, {Espigat}, {Farnier}, {Feinstein},
  {Fiasson}, {F{\"o}rster}, {Fontaine}, {F{\"u}{\ss}ling}, {Gabici}, {Gallant},
  {G{\'e}rard}, {Giebels}, {Glicenstein}, {Gl{\"u}ck}, {Goret}, {Hauser},
  {Hauser}, {Heinz}, {Heinzelmann}, {Henri}, {Hermann}, {Hinton}, {Hoffmann},
  {Hofmann}, {Holleran}, {Hoppe}, {Horns}, {Jacholkowska}, {de Jager}, {Jung},
  {Katarzy{\'n}ski}, {Katz}, {Kaufmann}, {Kendziorra}, {Kerschhaggl},
  {Khangulyan}, {Kh{\'e}lifi}, {Keogh}, {Komin}, {Kosack}, {Lamanna}, {Lenain},
  {Lohse}, {Marandon}, {Martin}, {Martineau-Huynh}, {Marcowith}, {Maurin},
  {McComb}, {Medina}, {Moderski}, {Moulin}, {Naumann-Godo}, {de Naurois},
  {Nedbal}, {Nekrassov}, {Niemiec}, {Nolan}, {Ohm}, {Olive}, {de O{\~n}a
  Wilhelmi}, {Orford}, {Ostrowski}, {Panter}, {Paz Arribas}, {Pedaletti},
  {Pelletier}, {Petrucci}, {Pita}, {P{\"u}hlhofer}, {Punch}, {Quirrenbach},
  {Raubenheimer}, {Raue}, {Rayner}, {Renaud}, {Reimer}, {Rieger}, {Ripken},
  {Rob}, {Rosier-Lees}, {Rowell}, {Rudak}, {Rulten}, {Ruppel}, {Sahakian},
  {Santangelo}, {Schlickeiser}, {Sch{\"o}ck}, {Schr{\"o}der}, {Schwanke},
  {Schwarzburg}, {Schwemmer}, {Shalchi}, {Skilton}, {Sol}, {Spangler},
  {Stawarz}, {Steenkamp}, {Stegmann}, {Superina}, {Tam}, {Tavernet}, {Terrier},
  {Tibolla}, {van Eldik}, {Vasileiadis}, {Venter}, {Venter}, {Vialle},
  {Vincent}, {Vivier}, {V{\"o}lk}, {Volpe}, {Wagner}, {Ward}, {Zdziarski}, \&
  {Zech}}]{HESS}
{Aharonian}, F., {Akhperjanian}, A.~G., {Anton}, G., {et~al.} 2009, Astronomy
  and Astrophysics, 499, 723, \dodoi{10.1051/0004-6361/200811357}

\bibitem[{{Aharonian}(2004)}]{aharonianbook}
{Aharonian}, F.~A. 2004, {Very high energy cosmic gamma radiation : a crucial
  window on the extreme Universe}, \dodoi{10.1142/4657}

\bibitem[{{Albert} {et~al.}(2020){Albert}, {Alfaro}, {Alvarez}, {Camacho},
  {Arteaga-Vel{\'a}zquez}, {Arunbabu}, {Avila Rojas}, {Ayala Solares},
  {Baghmanyan}, {Belmont-Moreno}, {BenZvi}, {Brisbois}, {Caballero-Mora},
  {Capistr{\'a}n}, {Carrami{\~n}ana}, {Casanova}, {Cotti}, {Couti{\~n}o de
  Le{\'o}n}, {De la Fuente}, {Diaz Hernandez}, {Diaz-Cruz}, {Dingus},
  {DuVernois}, {Durocher}, {D{\'\i}az-V{\'e}lez}, {Ellsworth}, {Engel},
  {Espinoza}, {Fan}, {Fang}, {Alonso}, {Fleischhack}, {Fraija},
  {Galv{\'a}n-G{\'a}mez}, {Garcia}, {Garc{\'\i}a-Gonz{\'a}lez}, {Garfias},
  {Giacinti}, {Gonz{\'a}lez}, {Goodman}, {Harding}, {Hernandez}, {Hinton},
  {Hona}, {Huang}, {Hueyotl-Zahuantitla}, {H{\"u}ntemeyer}, {Iriarte},
  {Jardin-Blicq}, {Joshi}, {Kieda}, {Lara}, {Lee}, {Le{\'o}n Vargas},
  {Linnemann}, {Longinotti}, {Luis-Raya}, {Lundeen}, {L{\'o}pez-Coto},
  {Malone}, {Marandon}, {Martinez}, {Martinez-Castellanos},
  {Mart{\'\i}nez-Castro}, {Matthews}, {Miranda-Romagnoli}, {Morales-Soto},
  {Moreno}, {Mostaf{\'a}}, {Nayerhoda}, {Nellen}, {Newbold}, {Nisa},
  {Noriega-Papaqui}, {Olivera-Nieto}, {Omodei}, {Peisker}, {P{\'e}rez Araujo},
  {P{\'e}rez-P{\'e}rez}, {Ren}, {Rho}, {Rivi{\`e}re}, {Rosa-Gonz{\'a}lez},
  {Ruiz-Velasco}, {Salazar}, {Salesa Greus}, {Sandoval}, {Schneider},
  {Schoorlemmer}, {Serna}, {Sinnis}, {Smith}, {Springer}, {Surajbali},
  {Tollefson}, {Torres}, {Torres-Escobedo}, {Ukwatta}, {Ure{\~n}a-Mena},
  {Weisgarber}, {Werner}, {Willox}, {Zepeda}, {Zhou}, {de Le{\'o}n},
  {{\'A}lvarez}, \& {HAWC Collaboration}}]{3hwc}
{Albert}, A., {Alfaro}, R., {Alvarez}, C., {et~al.} 2020, \apj, 905, 76,
  \dodoi{10.3847/1538-4357/abc2d8}

\bibitem[{Albert {et~al.}(2021)Albert, Alfaro, Alvarez, {\'{A}}lvarez, Camacho,
  Arteaga-Vel{\'{a}}zquez, Arunbabu, Rojas, Solares, Baghmanyan,
  Belmont-Moreno, BenZvi, Brisbois, Caballero-Mora, Capistr{\'{a}}n,
  Carrami{\~{n}}ana, Casanova, Cotti, Cotzomi, de~Le{\'{o}}n, la~Fuente,
  de~Le{\'{o}}n, Hernandez, Dingus, DuVernois, Durocher,
  D{\'{\i}}az-V{\'{e}}lez, Ellsworth, Engel, Espinoza, Fan, Alonso, Fraija,
  Galv{\'{a}}n-G{\'{a}}mez, Garc{\'{\i}}a-Gonz{\'{a}}lez, Garfias, Giacinti,
  Gonz{\'{a}}lez, Goodman, Harding, Hernandez, Hona, Huang,
  Hueyotl-Zahuantitla, Hüntemeyer, Iriarte, Jardin-Blicq, Joshi, Kieda, Lara,
  Lee, Lee, Vargas, Linnemann, Longinotti, Luis-Raya, Lundeen, Malone,
  Marandon, Martinez, Mart{\'{\i}}nez-Castro, Matthews, Miranda-Romagnoli,
  Morales-Soto, Moreno, Mostaf{\'{a}}, Nayerhoda, Nellen, Newbold, Nisa,
  Noriega-Papaqui, Olivera-Nieto, Omodei, Peisker, Araujo,
  P{\'{e}}rez-P{\'{e}}rez, Rho, Roh, Rosa-Gonz{\'{a}}lez, Ruiz-Velasco,
  Salazar, Greus, Sandoval, Schneider, Schoorlemmer, Serna-Franco, Smith,
  Springer, Surajbali, Tanner, Tollefson, Torres, Torres-Escobedo, Turner,
  Ure{\~{n}}a-Mena, Villase{\~{n}}or, Weisgarber, Willox, \& Zhou}]{2021uhe}
Albert, A., Alfaro, R., Alvarez, C., {et~al.} 2021, 911, L27,
  \dodoi{10.3847/2041-8213/abf4dc}

\bibitem[{{Albert} {et~al.}(2021){Albert}, {Alfaro}, {Alvarez},
  {Arteaga-Vel{\'a}zquez}, {Arunbabu}, {Avila Rojas}, {Ayala Solares},
  {Baghmanyan}, {Belmont-Moreno}, {Brisbois}, {Caballero-Mora},
  {Capistr{\'a}n}, {Carrami{\~n}ana}, {Casanova}, {Cotzomi}, {Couti{\~n}o de
  Le{\'o}n}, {De la Fuente}, {Diaz Hernandez}, {Dingus}, {DuVernois},
  {Durocher}, {Engel}, {Espinoza}, {Fraija}, {Garcia},
  {Garc{\'\i}a-Gonz{\'a}lez}, {Giacinti}, {Gonz{\'a}lez}, {Goodman}, {Harding},
  {Hinton}, {Hona}, {Huang}, {Hueyotl-Zahuantitla}, {Huentemeyer},
  {Jardin-Blicq}, {Joshi}, {Lee}, {Le{\'o}n Vargas}, {Linnemann}, {Longinotti},
  {Luis-Raya}, {Lundeen}, {L{\'o}pez-Coto}, {Malone}, {Martinez},
  {Mart{\'\i}nez-Castro}, {Matthews}, {Miranda-Romagnoli}, {Morales-Soto},
  {Moreno}, {Mostaf{\'a}}, {Nayerhoda}, {Nellen}, {Newbold}, {Nisa},
  {Noriega-Papaqui}, {Olivera-Nieto}, {Omodei}, {Peisker}, {P{\'e}rez Araujo},
  {P{\'e}rez-P{\'e}rez}, {Rho}, {Rosa-Gonz{\'a}lez}, {Ruiz-Velasco}, {Salazar},
  {Salesa Greus}, {Sandoval}, {Schneider}, {Schoorlemmer}, {Serna-Franco},
  {Smith}, {Springer}, {Surajbali}, {Tollefson}, {Torres}, {Turner},
  {Ure{\~n}a-Mena}, {Weisgarber}, {Willox}, {Zhou}, {de Le{\'o}n}, \& {HAWC
  Collaboration}}]{j2019}
{Albert}, A., {Alfaro}, R., {Alvarez}, C., {et~al.} 2021, \apj, 911, 143,
  \dodoi{10.3847/1538-4357/abecda}

\bibitem[{{Aliu} {et~al.}(2014){Aliu}, {Archambault}, {Aune}, {Behera},
  {Beilicke}, {Benbow}, {Berger}, {Bird}, {Buckley}, {Bugaev}, {Cardenzana},
  {Cerruti}, {Chen}, {Ciupik}, {Collins-Hughes}, {Connolly}, {Cui}, {Dumm},
  {Dwarkadas}, {Errando}, {Falcone}, {Federici}, {Feng}, {Finley},
  {Fleischhack}, {Fortin}, {Fortson}, {Furniss}, {Galante}, {Gall},
  {Gillanders}, {Griffin}, {Griffiths}, {Grube}, {Gyuk}, {Hanna}, {Holder},
  {Hughes}, {Humensky}, {Kaaret}, {Kertzman}, {Khassen}, {Kieda}, {Krennrich},
  {Kumar}, {Lang}, {Madhavan}, {Maier}, {McCann}, {Meagher}, {Millis},
  {Moriarty}, {Mukherjee}, {Nieto}, {O'Faol{\'a}in de Bhr{\'o}ithe}, {Ong},
  {Otte}, {Pandel}, {Park}, {Pohl}, {Popkow}, {Prokoph}, {Quinn}, {Ragan},
  {Rajotte}, {Ratliff}, {Reyes}, {Reynolds}, {Richards}, {Roache}, {Rousselle},
  {Sembroski}, {Shahinyan}, {Sheidaei}, {Smith}, {Staszak}, {Telezhinsky},
  {Tsurusaki}, {Tucci}, {Tyler}, {Varlotta}, {Vassiliev}, {Vincent}, {Wakely},
  {Ward}, {Weinstein}, {Welsing}, \& {Wilhelm}}]{VERITAS}
{Aliu}, E., {Archambault}, S., {Aune}, T., {et~al.} 2014, The Astrophysical
  Journal, 787, 166, \dodoi{10.1088/0004-637X/787/2/166}

\bibitem[{{Amato} {et~al.}(2003){Amato}, {Guetta}, \& {Blasi}}]{hadron1}
{Amato}, E., {Guetta}, D., \& {Blasi}, P. 2003, Astronomy and Astrophysics,
  402, 827, \dodoi{10.1051/0004-6361:20030279}

\bibitem[{{Bai} {et~al.}(2019){Bai}, {Bi}, {Bi}, {Cao}, {Chen}, {Chen},
  {Chiavassa}, {Cui}, {Dai}, {della Volpe}, {Di Girolamo}, {Di Sciascio},
  {Fan}, {Giacalone}, {Guo}, {He}, {He}, {Heller}, {Huang}, {Huang}, {Jia},
  {Ksenofontov}, {Leahy}, {Li}, {Li}, {Liang}, {Lipari}, {Liu}, {Liu}, {Liu},
  {Ma}, {Martineau-Huynh}, {Martraire}, {Montaruli}, {Ruffolo}, {Stenkin},
  {Su}, {Tam}, {Tang}, {Tian}, {Vallania}, {Vernetto}, {Vigorito}, {Wang},
  {Wang}, {Wang}, {Wang}, {Wang}, {Wang}, {Wei}, {Wei}, {Wu}, {Wu}, {Wu},
  {Yan}, {Yang}, {Yang}, {Yao}, {Yin}, {Yuan}, {Zhang}, {Zhang}, {Zhang},
  {Zhang}, {Zhang}, {Zhang}, {Zhao}, {Zhou}, {Zhu}, \& {Zhu}}]{lhaaso}
{Bai}, X., {Bi}, B.~Y., {Bi}, X.~J., {et~al.} 2019, arXiv e-prints,
  arXiv:1905.02773.
\newblock \doarXiv{1905.02773}

\bibitem[{{Bartoli} {et~al.}(2012){Bartoli}, {Bernardini}, {Bi}, {Bleve},
  {Bolognino}, {Branchini}, {Budano}, {Calabrese Melcarne}, {Camarri}, {Cao},
  {Cardarelli}, {Catalanotti}, {Cattaneo}, {Chen}, {Chen}, {Chen}, {Creti},
  {Cui}, {Dai}, {D'Al{\'\i} Staiti}, {Danzengluobu}, {Dattoli}, {De Mitri},
  {D'Ettorre Piazzoli}, {Di Girolamo}, {Ding}, {Di Sciascio}, {Feng}, {Feng},
  {Feng}, {Galeazzi}, {Giroletti}, {Gou}, {Guo}, {He}, {Hu}, {Hu}, {Huang},
  {Iacovacci}, {Iuppa}, {James}, {Jia}, {Labaciren}, {Li}, {Li}, {Li},
  {Liguori}, {Liu}, {Liu}, {Liu}, {Liu}, {Lu}, {Ma}, {Mancarella}, {Mari},
  {Marsella}, {Martello}, {Mastroianni}, {Montini}, {Ning}, {Pagliaro},
  {Panareo}, {Panico}, {Perrone}, {Pistilli}, {Qu}, {Ruggieri}, {Salvini},
  {Santonico}, {Shen}, {Sheng}, {Shi}, {Stanescu}, {Surdo}, {Tan}, {Vallania},
  {Vernetto}, {Vigorito}, {Wang}, {Wang}, {Wu}, {Wu}, {Xu}, {Xue}, {Yan},
  {Yang}, {Yang}, {Yao}, {Yuan}, {Zha}, {Zhang}, {Zhang}, {Zhang}, {Zhang},
  {Zhang}, {Zhang}, {Zhang}, {Zhaxiciren}, {Zhaxisangzhu}, {Zhou}, {Zhu},
  {Zhu}, {Zizzi}, \& {Argo-YBJ Collaboration}}]{ARGO}
{Bartoli}, B., {Bernardini}, P., {Bi}, X.~J., {et~al.} 2012, The Astrophysical
  Journal, 760, 110, \dodoi{10.1088/0004-637X/760/2/110}

\bibitem[{{Bednarek}(2003)}]{Bednarek}
{Bednarek}, W. 2003, Astronomy and Astrophysics, 407, 1,
  \dodoi{10.1051/0004-6361:20030929}

\bibitem[{{Breuhaus} {et~al.}(2021{\natexlab{a}}){Breuhaus}, {Hahn}, {Romoli},
  {Reville}, {Giacinti}, {Tuffs}, \& {Hinton}}]{breuhaus}
{Breuhaus}, M., {Hahn}, J., {Romoli}, C., {et~al.} 2021{\natexlab{a}}, \apjl,
  908, L49, \dodoi{10.3847/2041-8213/abe41a}

\bibitem[{{Breuhaus} {et~al.}(2021{\natexlab{b}}){Breuhaus}, {Reville}, \&
  {Hinton}}]{Breuhaus_et_al_2021_LHAASO}
{Breuhaus}, M., {Reville}, B., \& {Hinton}, J.~A. 2021{\natexlab{b}}, arXiv
  e-prints, arXiv:2109.05296.
\newblock \doarXiv{2109.05296}

\bibitem[{{Brisbois}(2021)}]{HAL}
{Brisbois}, C. 2021, in International Cosmic Ray Conference, Vol. 395, 37th
  International Cosmic Ray Conference (ICRC2021), 743

\bibitem[{{Brisbois}(2019)}]{chadsthesis}
{Brisbois}, C.~A. 2019, PhD thesis, Michigan Technological University

\bibitem[{Cao {et~al.}(2021)Cao, Aharonian, An, Axikegu, Bai, Bai,
  {et~al.}}]{lhaaso2}
Cao, Z., Aharonian, F., An, Q., {et~al.} 2021, Nature Astronomy, 594, 33,
  \dodoi{10.1038/s41586-021-03498-z}

\bibitem[{{Crestan} {et~al.}(2021){Crestan}, {Giuliani}, {Mereghetti},
  {Sidoli}, {Pintore}, \& {La Palombara}}]{2021MNRAS.505.2309C}
{Crestan}, S., {Giuliani}, A., {Mereghetti}, S., {et~al.} 2021, MRNAS, 505,
  2309, \dodoi{10.1093/mnras/stab1422}

\bibitem[{{Dame} {et~al.}(2001){Dame}, {Hartmann}, \& {Thaddeus}}]{DAME}
{Dame}, T.~M., {Hartmann}, D., \& {Thaddeus}, P. 2001, The Astrophysical
  Journal, 547, 792, \dodoi{10.1086/318388}

\bibitem[{{Di Palma} {et~al.}(2017){Di Palma}, {Guetta}, \& {Amato}}]{dipalma}
{Di Palma}, I., {Guetta}, D., \& {Amato}, E. 2017, Astrophysical Journal, 836,
  159, \dodoi{10.3847/1538-4357/836/2/159}

\bibitem[{{Downes} {et~al.}(1980){Downes}, {Pauls}, \& {Salter}}]{Downes}
{Downes}, A.~J.~B., {Pauls}, T., \& {Salter}, C.~J. 1980, \aap, 92, 47

\bibitem[{{Duvidovich} {et~al.}(2020){Duvidovich}, {Petriella}, \&
  {Giacani}}]{radio}
{Duvidovich}, L., {Petriella}, A., \& {Giacani}, E. 2020, \mnras, 491, 5732,
  \dodoi{10.1093/mnras/stz3414}

\bibitem[{{Evoli} {et~al.}(2018){Evoli}, {Linden}, \& {Morlino}}]{evoli}
{Evoli}, C., {Linden}, T., \& {Morlino}, G. 2018, Physical Review D, 98,
  063017, \dodoi{10.1103/PhysRevD.98.063017}

\bibitem[{{Giacalone} \& {Jokipii}(1999)}]{1999ApJ...520..204G}
{Giacalone}, J., \& {Jokipii}, J.~R. 1999, \apj, 520, 204,
  \dodoi{10.1086/307452}

\bibitem[{{Gonzalez-Garcia} {et~al.}(2009){Gonzalez-Garcia}, {Halzen}, \&
  {Mohapatra}}]{gonzalezgarcia}
{Gonzalez-Garcia}, M.~C., {Halzen}, F., \& {Mohapatra}, S. 2009, Astroparticle
  Physics, 31, 437, \dodoi{10.1016/j.astropartphys.2009.05.002}

\bibitem[{{Green}(2009)}]{GreenSNR}
{Green}, D.~A. 2009, Bulletin of the Astronomical Society of India, 37, 45.
\newblock \doarXiv{0905.3699}

\bibitem[{{Hahn}(2015)}]{gamera}
{Hahn}, J. 2015, in International Cosmic Ray Conference, Vol.~34, 34th
  International Cosmic Ray Conference (ICRC2015), 917

\bibitem[{{Halzen} {et~al.}(2017){Halzen}, {Kheirandish}, \&
  {Niro}}]{Halzen2017}
{Halzen}, F., {Kheirandish}, A., \& {Niro}, V. 2017, Astroparticle Physics, 86,
  46, \dodoi{10.1016/j.astropartphys.2016.11.004}

\bibitem[{{H.E.S.S. Collaboration}(2018)}]{vizier}
{H.E.S.S. Collaboration}. 2018, {VizieR Online Data Catalog: H.E.S.S. Galactic
  Plane Survey}, \dodoi{10.26093/cds/vizier.36120001}

\bibitem[{{HESS Collaboration} {et~al.}(2016){HESS Collaboration},
  {Abramowski}, {Aharonian}, {Benkhali}, {Akhperjanian}, {Ang{\"u}ner},
  {Backes}, {Balzer}, {Becherini}, {Tjus}, {Berge}, {Bernhard}, {Bernl{\"o}hr},
  {Birsin}, {Blackwell}, {B{\"o}ttcher}, {Boisson}, {Bolmont}, {Bordas},
  {Bregeon}, {Brun}, {Brun}, {Bryan}, {Bulik}, {Carr}, {Casanova},
  {Chakraborty}, {Chalme-Calvet}, {Chaves}, {Chen}, {Chr{\'e}tien},
  {Colafrancesco}, {Cologna}, {Conrad}, {Couturier}, {Cui}, {Davids},
  {Degrange}, {Deil}, {Dewilt}, {Djannati-Ata{\"\i}}, {Domainko}, {Donath},
  {Drury}, {Dubus}, {Dutson}, {Dyks}, {Dyrda}, {Edwards}, {Egberts}, {Eger},
  {Ernenwein}, {Espigat}, {Farnier}, {Fegan}, {Feinstein}, {Fernandes},
  {Fernand ez}, {Fiasson}, {Fontaine}, {F{\"o}rster}, {F{\"u}{\ss}ling},
  {Gabici}, {Gajdus}, {Gallant}, {Garrigoux}, {Giavitto}, {Giebels},
  {Glicenstein}, {Gottschall}, {Goyal}, {Grondin}, {Grudzi{\'n}ska}, {Hadasch},
  {H{\"a}ffner}, {Hahn}, {Hawkes}, {Heinzelmann}, {Henri}, {Hermann}, {Hervet},
  {Hillert}, {Hinton}, {Hofmann}, {Hofverberg}, {Hoischen}, {Holler}, {Horns},
  {Ivascenko}, {Jacholkowska}, {Jamrozy}, {Janiak}, {Jankowsky},
  {Jung-Richardt}, {Kastendieck}, {Katarzy{\'n}ski}, {Katz}, {Kerszberg},
  {Kh{\'e}lifi}, {Kieffer}, {Klepser}, {Klochkov}, {Klu{\'z}niak}, {Kolitzus},
  {Komin}, {Kosack}, {Krakau}, {Krayzel}, {Kr{\"u}ger}, {Laffon}, {Lamanna},
  {Lau}, {Lefaucheur}, {Lefranc}, {Lemi{\'e}re}, {Lemoine-Goumard}, {Lenain},
  {Lohse}, {Lopatin}, {Lu}, {Lui}, {Marand on}, {Marcowith}, {Mariaud}, {Marx},
  {Maurin}, {Maxted}, {Mayer}, {Meintjes}, {Menzler}, {Meyer}, {Mitchell},
  {Moderski}, {Mohamed}, {Mor{\r{a}}}, {Moulin}, {Murach}, {de Naurois},
  {Niemiec}, {Oakes}, {Odaka}, {{\"O}ttl}, {Ohm}, {Opitz}, {Ostrowski}, {Oya},
  {Panter}, {Parsons}, {Arribas}, {Pekeur}, {Pelletier}, {Petrucci}, {Peyaud},
  {Pita}, {Poon}, {Prokoph}, {P{\"u}hlhofer}, {Punch}, {Quirrenbach}, {Raab},
  {Reichardt}, {Reimer}, {Reimer}, {Renaud}, {de Los Reyes}, {Rieger},
  {Romoli}, {Rosier-Lees}, {Rowell}, {Rudak}, {Rulten}, {Sahakian}, {Salek},
  {Sanchez}, {Santangelo}, {Sasaki}, {Schlickeiser}, {Sch{\"u}ssler}, {Schulz},
  {Schwanke}, {Schwemmer}, {Seyffert}, {Simoni}, {Sol}, {Spanier}, {Spengler},
  {Spies}, {Stawarz}, {Steenkamp}, {Stegmann}, {Stinzing}, {Stycz}, {Sushch},
  {Tavernet}, {Tavernier}, {Taylor}, {Terrier}, {Tluczykont}, {Trichard},
  {Tuffs}, {Valerius}, {van der Walt}, {van Eldik}, {van Soelen},
  {Vasileiadis}, {Veh}, {Venter}, {Viana}, {Vincent}, {Vink}, {Voisin},
  {V{\"o}lk}, {Vuillaume}, {Wagner}, {Wagner}, {Wagner}, {Weidinger},
  {Weitzel}, {White}, {Wierzcholska}, {Willmann}, {W{\"o}rnlein}, {Wouters},
  {Yang}, {Zabalza}, {Zaborov}, {Zacharias}, {Zdziarski}, {Zech}, {Zefi}, \&
  {{\.Z}ywucka}}]{pevatron}
{HESS Collaboration}, {Abramowski}, A., {Aharonian}, F., {et~al.} 2016, Nature,
  531, 476, \dodoi{10.1038/nature17147}

\bibitem[{{HI4PI Collaboration} {et~al.}(2016){HI4PI Collaboration}, {Ben
  Bekhti}, {Fl{\"o}er}, {Keller}, {Kerp}, {Lenz}, {Winkel}, {Bailin},
  {Calabretta}, {Dedes}, {Ford}, {Gibson}, {Haud}, {Janowiecki}, {Kalberla},
  {Lockman}, {McClure-Griffiths}, {Murphy}, {Nakanishi}, {Pisano}, \&
  {Staveley-Smith}}]{HI4PI}
{HI4PI Collaboration}, {Ben Bekhti}, N., {Fl{\"o}er}, L., {et~al.} 2016,
  Astronomy and Astrophysics, 594, A116, \dodoi{10.1051/0004-6361/201629178}

\bibitem[{{Hobbs} {et~al.}(2004){Hobbs}, {Faulkner}, {Stairs}, {Camilo},
  {Manchester}, {Lyne}, {Kramer}, {D'Amico}, {Kaspi}, {Possenti}, {McLaughlin},
  {Lorimer}, {Burgay}, {Joshi}, \& {Crawford}}]{Parkes}
{Hobbs}, G., {Faulkner}, A., {Stairs}, I.~H., {et~al.} 2004, Monthly Notices of
  the Royal Astronomical Society, 352, 1439,
  \dodoi{10.1111/j.1365-2966.2004.08042.x}

\bibitem[{Kass \& Raftery(1995)}]{kass}
Kass, R.~E., \& Raftery, A.~E. 1995, Journal of the American Statistical
  Association, 90, 773, \dodoi{10.1080/01621459.1995.10476572}

\bibitem[{{Kierans}(2020)}]{kierans}
{Kierans}, C.~A. 2020, in Society of Photo-Optical Instrumentation Engineers
  (SPIE) Conference Series, Vol. 11444, Society of Photo-Optical
  Instrumentation Engineers (SPIE) Conference Series, 1144431,
  \dodoi{10.1117/12.2562352}

\bibitem[{{Kolmogorov}(1941)}]{kolmogorov}
{Kolmogorov}, A. 1941, Akademiia Nauk SSSR Doklady, 30, 301

\bibitem[{{Kostunin} {et~al.}(2021)}]{HESS2021ICRC}
{Kostunin}, D., {et~al.} 2021, in International Cosmic Ray Conference, Vol.
  395, 37th International Cosmic Ray Conference (ICRC2021), 779,
  \dodoi{https://doi.org/10.22323/1.395.0779}

\bibitem[{{Li} {et~al.}(2021){Li}, {Liu}, {de Ona Wilhelmi}, {Torres}, {Liu},
  {Kerr}, {Buehler}, {Su}, {He}, \& {Xiao}}]{fermiPWN}
{Li}, J., {Liu}, R.-Y., {de Ona Wilhelmi}, E., {et~al.} 2021, arXiv e-prints,
  arXiv:2102.05615.
\newblock \doarXiv{2102.05615}

\bibitem[{{Lyne} {et~al.}(2017){Lyne}, {Stappers}, {Bogdanov}, {Ferdman},
  {Freire}, {Kaspi}, {Knispel}, {Lynch}, {Allen}, {Brazier}, {Camilo},
  {Cardoso}, {Chatterjee}, {Cordes}, {Crawford}, {Deneva}, {Hessels}, {Jenet},
  {Lazarus}, {van Leeuwen}, {Lorimer}, {Madsen}, {McKee}, {McLaughlin},
  {Parent}, {Patel}, {Ransom}, {Scholz}, {Seymour}, {Siemens}, {Spitler},
  {Stairs}, {Stovall}, {Swiggum}, {Wharton}, {Zhu}, {Aulbert}, {Bock},
  {Eggenstein}, {Fehrmann}, \& {Machenschalk}}]{PAFLA}
{Lyne}, A.~G., {Stappers}, B.~W., {Bogdanov}, S., {et~al.} 2017, The
  Astrophysical Journal, 834, 137, \dodoi{10.3847/1538-4357/834/2/137}

\bibitem[{Manchester {et~al.}(2005)Manchester, Hobbs, Teoh, \&
  Hobbs}]{Manchester2005}
Manchester, R.~N., Hobbs, G.~B., Teoh, A., \& Hobbs, M. 2005, The Astronomical
  Journal, 129, 1993, \dodoi{10.1086/428488}

\bibitem[{{Moderski} {et~al.}(2005){Moderski}, {Sikora}, {Coppi}, \&
  {Aharonian}}]{moderski}
{Moderski}, R., {Sikora}, M., {Coppi}, P.~S., \& {Aharonian}, F. 2005, Monthly
  Notices of the Royal Astronomical Society, 363, 954,
  \dodoi{10.1111/j.1365-2966.2005.09494.x}

\bibitem[{{Pandel}(2015)}]{XMM}
{Pandel}, D. 2015, in International Cosmic Ray Conference, Vol.~34, 34th
  International Cosmic Ray Conference (ICRC2015), 743.
\newblock \doarXiv{1512.08140}

\bibitem[{{Particle Data Group} {et~al.}(2020){Particle Data Group}, Zyla,
  Barnett, Beringer, Dahl, Dwyer, Groom, Lin, Lugovsky, Pianori, Robinson,
  Wohl, Yao, Agashe, Aielli, Allanach, Amsler, Antonelli, Aschenauer, Asner,
  Baer, Banerjee, Baudis, Bauer, Beatty, Belousov, Bethke, Bettini, Biebel,
  Black, Blucher, Buchmuller, Burkert, Bychkov, Cahn, Carena, Ceccucci, Cerri,
  Chakraborty, Chivukula, Cowan, D'Ambrosio, Damour, de~Florian, de~Gouvêa,
  DeGrand, de~Jong, Dissertori, Dobrescu, D'Onofrio, Doser, Drees, Dreiner,
  Eerola, Egede, Eidelman, Ellis, Erler, Ezhela, Fetscher, Fields, Foster,
  Freitas, Gallagher, Garren, Gerber, Gerbier, Gershon, Gershtein, Gherghetta,
  Godizov, Gonzalez-Garcia, Goodman, Grab, Gritsan, Grojean, Grünewald, Gurtu,
  Gutsche, Haber, Hanhart, Hashimoto, Hayato, Hebecker, Heinemeyer, Heltsley,
  Hernández-Rey, Hikasa, Hisano, Höcker, Holder, Holtkamp, Huston, Hyodo,
  Johnson, Kado, Karliner, Katz, Kenzie, Khoze, Klein, Klempt, Kowalewski,
  Krauss, Kreps, Krusche, Kwon, Lahav, Laiho, Lellouch, Lesgourgues, Liddle,
  Ligeti, Lippmann, Liss, Littenberg, Lourengo, Lugovsky, Lusiani, Makida,
  Maltoni, Mannel, Manohar, Marciano, Masoni, Matthews, Meißner, Mikhasenko,
  Miller, Milstead, Mitchell, Mönig, Molaro, Moortgat, Moskovic, Nakamura,
  Narain, Nason, Navas, Neubert, Nevski, Nir, Olive, Patrignani, Peacock,
  Petcov, Petrov, Pich, Piepke, Pomarol, Profumo, Quadt, Rabbertz, Rademacker,
  Raffelt, Ramani, Ramsey-Musolf, Ratcliff, Richardson, Ringwald, Roesler,
  Rolli, Romaniouk, Rosenberg, Rosner, Rybka, Ryskin, Ryutin, Sakai, Salam,
  Sarkar, Sauli, Schneider, Scholberg, Schwartz, Schwiening, Scott, Sharma,
  Sharpe, Shutt, Silari, Sjöstrand, Skands, Skwarnicki, Smoot, Soffer, Sozzi,
  Spanier, Spiering, Stahl, Stone, Sumino, Sumiyoshi, Syphers, Takahashi,
  Tanabashi, Tanaka, Taševský, Terashi, Terning, Thoma, Thorne, Tiator,
  Titov, Tkachenko, Tovey, Trabelsi, Urquijo, Valencia, Van~de Water, Varelas,
  Venanzoni, Verde, Vincter, Vogel, Vogelsang, Vogt, Vorobyev, Wakely,
  Walkowiak, Walter, Wands, Wascko, Weinberg, Weinberg, White, Wiencke,
  Willocq, Woody, Workman, Yokoyama, Yoshida, Zanderighi, Zeller, Zenin, Zhu,
  Zhu, Zimmermann, Anderson, Basaglia, Lugovsky, Schaffner, \& Zheng}]{pdg2021}
{Particle Data Group}, Zyla, P.~A., Barnett, R.~M., {et~al.} 2020, Progress of
  Theoretical and Experimental Physics, 2020, \dodoi{10.1093/ptep/ptaa104}

\bibitem[{{Pinat} {et~al.}(2017){Pinat}, {Aguilar S{\'a}nchez}, \& {IceCube
  Collaboration}}]{IceCubeExtended}
{Pinat}, E., {Aguilar S{\'a}nchez}, J.~A., \& {IceCube Collaboration}. 2017, in
  International Cosmic Ray Conference, Vol. 301, 35th International Cosmic Ray
  Conference (ICRC2017), 963

\bibitem[{{Popescu} {et~al.}(2017){Popescu}, {Yang}, {Tuffs}, {Natale},
  {Rushton}, \& {Aharonian}}]{Popescu2017}
{Popescu}, C.~C., {Yang}, R., {Tuffs}, R.~J., {et~al.} 2017, \mnras, 470, 2539,
  \dodoi{10.1093/mnras/stx1282}

\bibitem[{Schwarz(1978)}]{schwarz1978}
Schwarz, G. 1978, Ann. Statist., 6, 461, \dodoi{10.1214/aos/1176344136}

\bibitem[{Smith(2015)}]{Smith2015}
Smith, A.~J. 2015, Proceedings of Science (34th International Cosmic Ray
  Conference), 966.
\newblock \url{https://pos.sissa.it/236/966/pdf}

\bibitem[{Sudoh {et~al.}(2021)Sudoh, Linden, \& Hooper}]{2021sudoh}
Sudoh, T., Linden, T., \& Hooper, D. 2021, 2021, 010,
  \dodoi{10.1088/1475-7516/2021/08/010}

\bibitem[{{Vianello} {et~al.}(2015){Vianello}, {Lauer}, {Younk}, {Tibaldo},
  {Burgess}, {Ayala}, {Harding}, {Hui}, {Omodei}, \& {Zhou}}]{threeml}
{Vianello}, G., {Lauer}, R.~J., {Younk}, P., {et~al.} 2015, arXiv e-prints,
  arXiv:1507.08343.
\newblock \doarXiv{1507.08343}

\bibitem[{Wilks(1938)}]{Wilks1938}
Wilks, S.~S. 1938, The Annals of Mathematical Statistics, 9, 60,
  \dodoi{10.1214/aoms/1177732360}

\bibitem[{{Zhou} {et~al.}(2017){Zhou}, {Rho}, {Vianello}, \& {HAWC
  Collaboration}}]{diffuseICRC}
{Zhou}, H., {Rho}, C.~D., {Vianello}, G., \& {HAWC Collaboration}. 2017, in
  International Cosmic Ray Conference, Vol. 301, 35th International Cosmic Ray
  Conference (ICRC2017), 689.
\newblock \doarXiv{1709.03619}

\end{thebibliography}

\end{document}